\documentclass[11pt]{article}
\usepackage{graphicx}  
\usepackage{float}
\usepackage{amssymb}
\usepackage{amsmath}
\usepackage{amsfonts}
\usepackage{mathrsfs}
\usepackage[scaled=.92]{helvet}
\usepackage{times}

\usepackage{parskip}

\newtheorem{theorem}{Theorem}[section]
\newtheorem{lemma}[theorem]{Lemma}

\newtheorem{corollary}[theorem]{Corollary}
\newtheorem{alg}[theorem]{Algorithm}

\newtheorem{remark}[theorem]{Remark}

\def\qed{\hfil {\vrule height5pt width2pt depth2pt}}
\def\S{\mathcal{S}}
\def\C{\mathcal{C}}

\def\wrt{w.r.t.\,}

\def\B{{\mathbf{B}}}
\def\FB{{\mathbf{F}}}
\def\EB{{\mathbf{E}}}
\def\SB{{\mathbf{S}}}
\def\NB{{\mathbf{N}}}
\def\PB{{\mathbf{P}}}
\def\CB{{\mathbf{C}}}

\def\Res{\hbox{\rm{Res}}}
\def\res{\hbox{\rm{Res}}}
\def\sat{\hbox{\rm{Sat}}}

\def\bref#1{(\ref{#1})}

\def\Q{{\mathbb Q}}
\def\R{{\mathbb R}}

\newcommand{\intbox}{{\setlength{\unitlength}{.33mm}\framebox(4, 7){}\, }}
\def\IBQ{{\intbox{\mathbb Q}}}
\def\ROOTI{{\bf RootIsol}}

\def\PP{{\mathcal P}}
\def\PE{{\mathcal E}}

\def\PF{{\mathcal F}}

\def\EP{{\mathcal{EP}}}
\def\EE{{\mathcal{EE}}}
\def\EC{{\mathcal{EC}}}

\def\MB{\mathcal {B}}

\def\M{{\mathcal M}}
\def\mb{{\mathcal {M}\mathcal {B}}}
\def\Mp{{\mathcal {M}\mathcal {P}}}
\def\me{{\mathcal {M}\mathcal {E}}}
\def\mf{{\mathcal {M}\mathcal {F}}}
\def\mp{{\mathcal {M}\mathcal {P}}}

\def\QQ{{\mathcal Q}}
\def\TT{{\mathcal T}}
\def\LL{{\mathcal L}}
\def\RR{{\mathcal R}}

\def\SP{{\mathcal{SP}}}
\def\SE{{\mathcal{SE}}}
\def\SF{{\mathcal{SF}}}

\def\SBX{{\mathcal{SB}}}

\def\MM{{\mathcal M}}

\def\G{{\mathscr{G}}}
\def\EG{\mathcal{EG}}
\def\P{{\mathscr{P}}}

\def\XB{{\mathcal{X}}}
\def\YB{{\mathcal{Y}}}
\def\ZB{{\mathcal{Z}}}

\title{Ambient Isotopic Meshing of Implicit Algebraic Surface with Singularities}

\author{Jin-San Cheng$^{1,2}$\thanks{e-mail: jcheng@amss.ac.cn}, \hspace{1cm}
 Xiao-Shan Gao$^1$\thanks{e-mail:xgao@mmrc.iss.ac.cn}, \hspace{1cm} Jia Li$^1$\thanks{lijia@mmrc.iss.ac.cn}\\
$^1$Key Lab of Mathematics Mechanization\\ Institute of Systems
Science, AMSS, Academia Sinica\\ $^2$Loria, INRIA Nancy}

\date{}

\begin{document}

\maketitle


\begin{abstract}
A complete method is proposed to compute a certified, or ambient
isotopic, meshing for an implicit algebraic surface with
singularities. By certified, we mean a meshing with correct topology
and any given geometric precision. We propose a symbolic-numeric
method to compute a certified meshing for the surface inside a box
containing singularities and use a modified Plantinga-Vegter
marching cube method to compute a certified meshing for the surface
inside a box without singularities. Nontrivial examples are given to
show the effectiveness of the algorithm (see Fig. \ref{fig-first}).
To our knowledge, this is the first method to compute a certified
meshing for surfaces with singularities.

\end{abstract}

\section{Introduction}
\begin{figure}[ht]
\centering
\begin{minipage}{0.9\textwidth}
   \includegraphics[width=1.01in]{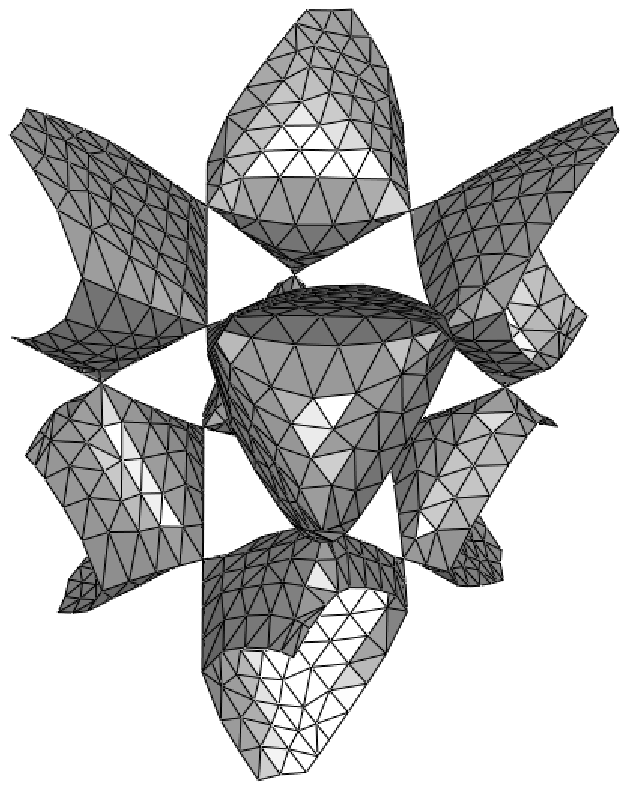}
   \includegraphics[width=1.08in]{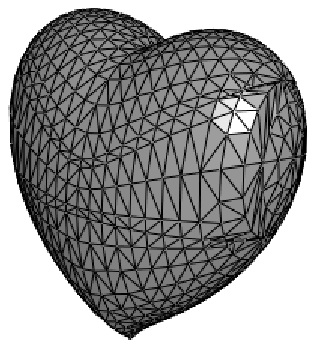}
   \includegraphics[width=1.02in]{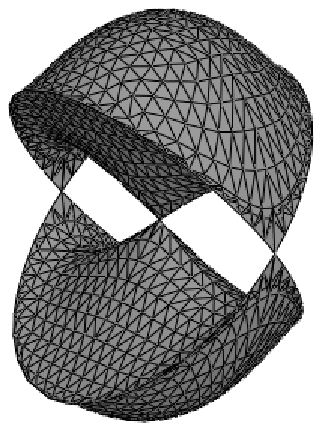}
   \includegraphics[width=1.18in]{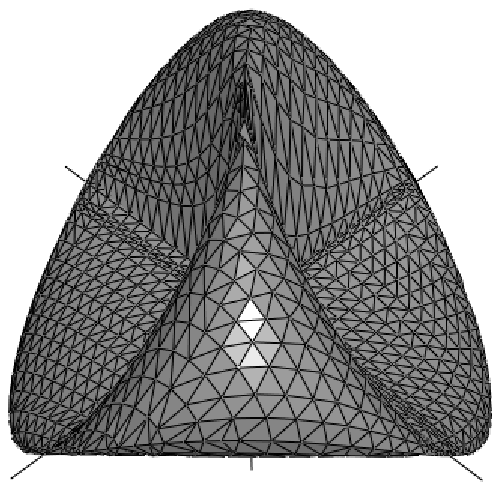}
   \includegraphics[width=0.9in]{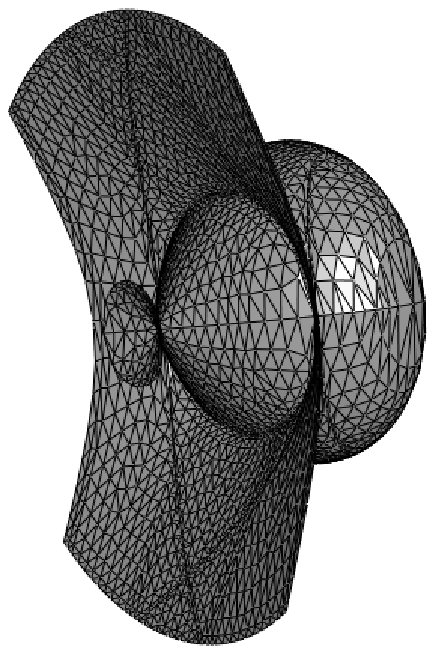}
\caption{Isotopic meshing for surfaces with singular points and
singular curves} \label{fig-first}
\end{minipage}
\end{figure}

To determine the topology of a given algebraic surface and to use
triangular meshes to approximately represent the surface are
fundamental operations in computer graphics and geometric modeling.
Meshing of surfaces could be used to display the surface correctly
and to perform engineering applications on the surface, such as the
finite element analysis. A survey on this topic can be found in
\cite{bo0}.

We consider an implicit surface defined by $f(x, y, z)=0$ where
$f(x,y,z)$ is a square free polynomial with rational numbers as
coefficients. There exists a large amount of work on meshing
implicit surfaces. Please see the work \cite{all1,baj1,blo1} and the
literatures cited in them.
Recent work focuses on isotopic meshing  \cite{bo0}. Simply
speaking, a meshing is called {\bf isotopic} if it has the same
topology and the same geometry as the surface (for definition see
Section 2). A meshing is called {\bf ambient isotopic} or {\bf
certified} if it is isotopic and approximates the surface to any
give precision.
There exist four main approaches to compute isotopic meshings for
surfaces: the marching cube method, the Morse theory method, the
Delaunay refinement method, and the CAD (Cylindrical Algebraic
Decomposition) based method.

The famous marching cube method repeatedly subdivides the space into
smaller cubes until the structure of the surface inside each cube is
known \cite{wil}. For implicit surfaces, Snyder proposed the
{globally parameterizable} criterion for that purpose \cite{snyder}.
Plantinga and Vegter proposed the small normal variation condition
which leads to a better meshing algorithm \cite{pv1,pv2}.

Hart et al proposed a method based on Morse theory
\cite{hart1,hart2,hart3}. The idea is to check when the topology of
$f(x,y,z)=a$ will change for a parameter $a$. When $a$ changes from
some initial value where $f(x,y,z)=a$ has no solution to $a=0$, the
topology of the surface is found.
Fortuna et al presented improved algorithms for surfaces in the
projective space \cite{for1,for2}.

For a set of points on the surface, one can form the restricted
Delaunay triangles and the corresponding Delaunay triangulation can
be used to approximate the surface. Boissonnat and Oudot proved that
when the sample point set satisfies certain conditions, the Delaunay
triangulation has the same topology as the surface \cite{bo1,bo2}.
Cheng et al established similar results using different strategies
\cite{cheng1}.

The CAD method proposed by Collins can be used to divide the
Euclidean space into cylindrical cells such that the given surface
has the same sign on each of the cells. Then to determine the
topology of the surface, we need only give the adjacency information
between the cells \cite{cad3,mcc}.
Alone this line, new ideas are introduced to compute the topology of
surfaces \cite{chengtop,mou}.
%
%

All the above methods except the one based on CAD  work for surfaces
without singularities only. In this paper, we  give a method to
compute a {certified} meshing for implicit algebraic surfaces with
singularities.
The method is a hybrid one based on the CAD approach and the
marching cube approach. We propose a CAD based method to compute a
certified meshing for the surface inside a box containing
singularities and use a modified Plantinga-Vegter method to compute
a certified meshing for the surface inside a box without
singularities.
Our main contribution is how to treat the singularities.

This paper consists of three parts. The algorithms for surfaces are
the main contributions.
In Section 3,  a new method is proposed to compute a certified
meshing for a plane algebraic curve. This section also provides
preparations for algorithms about surfaces.
There exist many methods to compute the topology of plane curves,
e.g., \cite{cad2,yap-issac2008,eigenwillig,gon1}.
Our contribution is to give an interval based method to compute the
adjacency information and to give an ambient isotopic meshing for a
curve. The method in \cite{yap-issac2008} can also compute an
ambient isotopic meshing based on root bounds of equation systems.
Our method is based on symbolic-numerical computation, which is
practically more effective.

In Section 4, a new method is proposed to compute an isotopic
meshing for a surface. The method has two advantages. First, we use
symbolic computation methods to guarantee the completeness and
whenever possible use interval arithmetics to increase the
efficiency. Actually, computations of algebraic numbers are totally
avoided. The work \cite{cad2,mcc} uses algebraic numbers.
Second, our algorithm does not change the surface to generic
positions as done in \cite{mou}, which is generally expensive.
Our method need only to project the surface once, while the
algorithm proposed in \cite{mou} need to do projections twice.

In Section 5, a method is proposed to compute a certified meshing
for a surface.
A well-known technical to treat a singular point $P$ is to find a
{\bf segregating box} which contains $P$ but does not intersect the
surface at its bottom and top faces. We f extend this concept to
singular curve segments and give an interval based method to compute
such boxes and meshes in the boxes.
Another key ingredient is a careful analysis of the extremal points
of surfaces and spatial curves.
It is pointed out in \cite{bo0}, that the method in \cite{mou}
``makes no guarantees about the geometric accuracy of the mesh, and
it cannot be extended in a straightforward way to provide a more
accurate mesh." To our knowledge, the method proposed in this paper
is the first one to compute a certified meshing for surfaces with
singularities.

Algorithms in Sections 3 and 4 are implemented in Maple and
nontrivial examples are used to show that the algorithm is quite
effective for surfaces with singular points and curves.

\section{Preparations}
In this section, we give several known results and algorithms needed
in this paper. Following \cite{bo0}, we will compute a meshing with
correct topology for a curve or a surface in the following sense.

An {\bf isotopic meshing} for a variety $\S\subset\R^n\, (n=2,3)$
consists of a graph/polyhedron $\G$ (for $n=2,3$) and a continuous
mapping $\gamma: \R^n\times[0,1] \rightarrow \R^n$
%
 which, for any fixed $t\in[0,1]$, is a homeomorphism
$\gamma(\cdot,t)$ from $R^n$ to itself, and which continuously
deforms $\G$ into $S$: $\gamma(\cdot,0)=id$, $\gamma(\G,1)=\S$.

For a number $\epsilon>0$, an {\bf $\epsilon$-meshing} for $\S$ is
an isotopic meshing $\G$ for $\S$, which gives an
$\epsilon$-approximation for $S$ in the following sense $
\parallel P-\gamma(P,1) \parallel\le \epsilon \hbox{ for all } P\in
 \G.$
Please note that isotopy is stronger than  homeomorphism \cite{bo0}.

\subsection{Real root isolation of triangular system}

A basic step of our algorithm  is to isolate the real roots of a
{\bf triangular  system} which consists of equations like
 \begin{equation}\label{eq-tn}
 \Sigma_n=\{f_1(x_1), f_2(x_1, x_2), \ldots, f_n(x_1, x_2, \ldots, x_n)\}
 \end{equation}
where $f_i\in \Q[x_1, \ldots, x_i]$ involves $x_i$ effectively.

We use intervals to isolate real numbers: let $\intbox \Q$ denote
the set of intervals of the form $[a, b]$ where $a< b\in \Q$. The
{\bf length} of an interval box $\B_n=
[a_1,b_1]\times\cdots\times[a_n, b_n]\in\IBQ^n$ is defined to be
$|\B_n| = \max_i (b_i-a_i)$.

In this paper, when we say a {\bf point}, we mean  a point
$P=(\xi_1, \ldots, \xi_n)$ with real algebraic numbers as
coordinates, which is represented by a triangular system $\Sigma_n$
like \bref{eq-tn} with $P$ as a solution and an isolation box $\B_n$
for $\xi$.
For instance $\sqrt{2}$ is represented by $x_1^2-2=0$ and $(1,2)$.

Now,  we give a formal description of the root isolation algorithm.
\begin{alg}\label{alg-rootiso}
\ROOTI$(\Sigma_n, \B_n, \epsilon)$. \quad The input consists of a
triangular system $\Sigma_n$ of form \bref{eq-tn},  a box
$\B_n\in\IBQ^n$,  and a positive number $\epsilon$. The output is a
set of isolation boxes for all the real roots of $\Sigma_n=0$ in
$\B_n$ such that the length of the isolation boxes is smaller than
$\epsilon$ and any two of the isolation boxes are disjoints.
\end{alg}

A modified version of the root isolation algorithms in
\cite{chengroots,Rouillier99} is used in our implementation.

Let $f(x_1, \ldots, x_n)\in\Q[x_1, \ldots, x_n]$ and $\B_n=
[a_1,b_1]\times\cdots\times[a_n, b_n]\in\IBQ^n$.
The {\bf box operation} $\intbox f(\B_n)$ returns an interval
containing all the points $\{f(x_1, \ldots, x_n)\,|\, a_i\le x_i\le
b_i,i=1,\ldots,n\}$. Furthermore, when $|\B_n|$ approaches to zero,
the length of interval $\intbox f(\B_n)$ also approaches to zero.
If $a_i>0$ and $b_i>0$, we can construct  $\intbox f(\B_n)$ as
follows

$$ \intbox f(\B_n)=f^+(b_1,\ldots,b_n)-f^-(a_1,\ldots,a_n)$$
where $f = f^+ - f^-$ such that  $f^+, f^-\in \Q[x_1\ldots,x_n]$
each has only positive coefficients and minimal number of monomials.
For the general case, please consult \cite{chengroots}.
It is clear that such an operation satisfies the following property.
%

\begin{lemma}\label{cor-sl} If $\xi=(\xi_1, \ldots, \xi_n)$ is not a zero of
$f(x_1,\ldots,x_n)=0$ and $\B_n$ an isolation box for $\xi$. Then if
the length of $\B_n$ is small enough, the interval $\intbox f(\B_n)$
will not contain $(0,\ldots,0)$, which means that $\B_n$ has no
intersections with $f=0$. We denote this as $\intbox f(\B_n)\ne0$.
\end{lemma}

\subsection{Delineable polynomials}

Delineable polynomials are important in determining the topology of
algebraic surfaces. Let $f(x_1$, $\ldots, x_{r-1}, x_r)\in
\mathbb{R}[x_1, \ldots, x_r]$ and $P=(p_1, \ldots, p_r)$ a point of
$\mathbb{R}^r$. We say that $f$ has {\bf order} $k$ at point $P$,
%
%
if $k\ge0$ is the least non-negative integer such that some partial
derivative of total order $k$ does not vanish at $P$. And $f$ is
said to be {\bf order-invariant} in a subset $R$ of $\mathbb{R}^r$
provided that the order of $f$ is the same at every point of $R$.

For simplification, we denote the $(r-1)$-tuple $(x_1, \ldots,
x_{r-1})$ as $\bar{x}$. An $r$-variate polynomial $f(\bar{x},  x_r)$
over the reals is said to be {\bf  delineable} on a submanifold $R$
of $\mathbb{R}^{r-1}$ if it holds that:
\begin{description}
\item[(1)]  the portion of the real variety of $f$ that lies in the
cylinder $R\times\mathbb{R}$ over $R$ consists of the union of the
function graphs of some $k>0$ analytic functions
$\theta_1<\ldots<\theta_k$ from $R$ into $\mathbb{R}$; and
\item[(2)]
there exist positive integers $m_i$ such that for every $\alpha\in
R$,  the multiplicity of the root of $f(\alpha, x_r)$ corresponding
to $\theta_i$ is $m_i$.
\end{description}

Polynomial $f$ is said to {\bf vanish identically} on $R$ if $f(P,
x_r)=0$ for every point $P\in R$. In addition, $f$ is said to be
{\bf degree-invariant} on $R$ if the degree of $f(P, x_r)$ as a
polynomial in $x_r$ is the same for every point $P\in R$.
In this situation, the following theorem holds (see \cite{mcc}, pp.
246).
\begin{theorem}\label{th-mcc}(McCallum and Collins)
Let $f(\bar{x}, x_r)$ be  a polynomial in $\mathbb{R}[\bar{x}, x_r]$
of positive degree in $x_r$. Let $D(\bar{x})$ be the discriminant of
$f$ as a univariate polynomial in $x_r$ and suppose that
$D(\bar{x})$ is a nonzero polynomial. Let $R$ be a connected
submanifold of $\mathbb{R}^{r-1}$ on which $f$ is degree-invariant
and does not vanish identically,  and over which $D$ is
order-invariant. Then,  $f$ is  delineable on $R$. 
%
\end{theorem}

The following theorem improves the above result.
\begin{theorem}\label{th-brown}(\cite{brown})
Let $f\in\mathbb{R}(\bar{x}, x_r)$ ($r\ge2$) be an $r$-variate
polynomial  of positive degree in $x_r$ with discriminant
$D(\bar{x})\neq0$. Let $R$ be a connected submanifold of
$\mathbb{R}^{r-1}$ in which $D$ is order-invariant, the leading
coefficient of $f$ \wrt  $x_r$ is sign-invariant, and such that $f$
vanishes identically at no point in $R$. Then, $f$ is
degree-invariant on $R$. 
%
\end{theorem}

\section{Ambient isotopic meshing of plane  curve}
In this section, we give an algorithm to compute an isotopic meshing
for an algebraic curve.
The main purpose of this section is to provide preliminary
algorithms for later sections.
We also give a new and fast method to compute the adjacency
information based on interval arithmetics.

\subsection{Determine the topology of plane algebraic curve}
\label{sec-c1}

We use a graph to represent the topology of a plane curve.
A {\bf topology graph} is a graph $\G= \{\PP, \PE\}$ where
\begin{itemize}
\item $\PP$ is a set of plane points defined by triangular systems
$\Sigma_i$ and isolation boxes $\B_{i,j}$: {\small
\begin{eqnarray}\label{eq-ps2}
 \PP &=& \{P_{i, j}=(\alpha_i,  \beta_{i, j}),  0\le i\le s,  0\le j \le s_i
 \}\\
  \Sigma_i &=& \{h_i(x), g_i(x, y)\}, \B_{i,j} = [a_i,b_i]\times[c_{i,
 j},d_{i,j}]\nonumber
 \end{eqnarray}}
where $\alpha_0<\alpha_1<\cdots<\alpha_s$ and $\beta_{i,
0}<\beta_{i, 1}<\cdots<\beta_{i, s_i}$.
When drawing the graph, we use %
$M_{i,j} = ((a_i+b_i)/2,(c_{i, j}+d_{i, j})/2)$ to represent
$P_{i,j}$.

\item $\PE=\{(P_1, P_2) | P_1, P_2\in\PP, $ such that either
$P_1=P_{i, p}, P_2=P_{i+1, q}$ \hbox{ or } $P_1=P_{i, p}, P_2=P_{i,
p+1}\}$.
In the first case,  the edge is called {\bf non-vertical}. In the
second case,  the edge is called {\bf $x$-vertical}. We further
assume that any two edges do not intersect except at the end points.
\end{itemize}

Consider  a plane  algebraic  curve $\C: g(x, y)=0$
%
%
where $g(x, y)\in \mathbb{Q}[x, y]$ is a square free polynomial.
A point $P_0$ is  an {\bf $x$-critical point} of $\C$ if
$g(P_0)=g_y(P_0)=0$.

We will consider the part of $\mathcal{C}$ in a bounding box
 \begin{equation}\label{eq-b2}\B_2=[\XB_1,\XB_2]\times[\YB_1,\YB_2]\in\IBQ^2\end{equation}
which is denoted as $\mathcal{C}_{\B_2}=\C\cap\B_2$. In the rest
part of this paper, $\B_2$ is always assumed to be of this form.

Let $P$ be a point on curve $\C$,  the {\bf left (right) branch
number of $P$}, also denoted as $L\#(P)$ ($R\#(P)$),  is the number
of curve segments of $\C$ which pass through  $P$ and are on the
left (right) side of $P$ in a small neighborhood of $P$.
%

\begin{figure}[ht]
\centering
\includegraphics[scale=0.35]{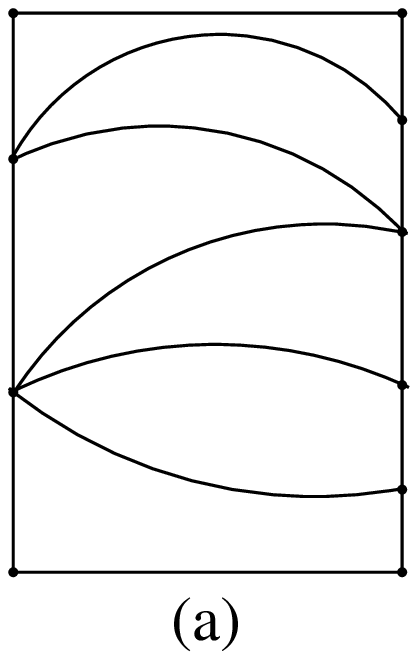}\quad
\includegraphics[scale=0.35]{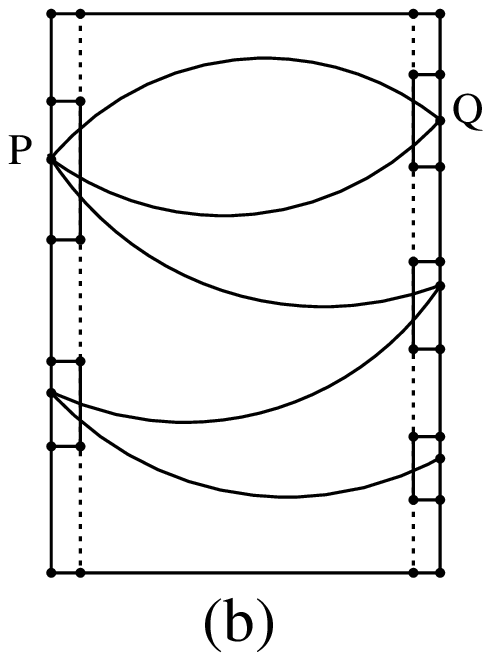}\quad
\includegraphics[scale=0.35]{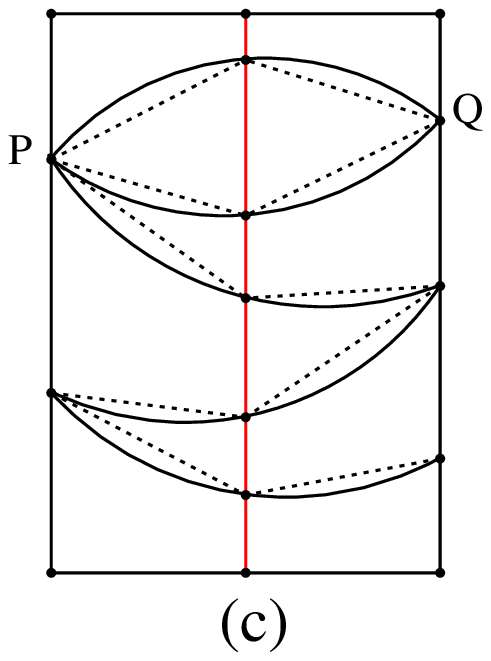}
\caption{Curve segments inside a segregating box.}\label{fig-cur1}
\end{figure}

We introduce the key concept of segregating box.
A box $\B=[a, b]\times[c,d]\in\IBQ^2$ is called {\bf segregating}
\wrt $\C$ if
$$\C\cap[a,b]\times[c,c]=\C\cap[a,b]\times[d,d]=\emptyset.$$
A curve behaves nicely in a segregating box, as illustrated by the
following lemma. See Fig. \ref{fig-cur1}(a) for an illustration.

\begin{lemma}\label{prop-22}
Let  $\B=[a, b]\times[c, d]\in\IBQ^2$ be a box segregating \wrt $\C$
and the interior of $\B$ contains no $x-$critical points of $\C$.
Let $\C$ intersect the left and right boundaries of $\B$ at points
$L_i, i=1,\ldots, l$ and $R_j, j=1,\ldots,r$ respectively.
Then $\C$ is delineable over $R=(a,b)$ and the number of curve
segments of $\C$ inside $\B$ equals $\sum_{i=1}^l R\#(L_i) =
\sum_{j=1}^r R\#(R_i)$. (See Figure \ref{fig-cur1}(a) for an
illustration)
\end{lemma}
%
{\em Proof.}
Note that the leading coefficient $C(x)$ of $g(x,y)$ \wrt  to $y$ is
a factor of the discriminant $D(x)$ of $g(x,y)$ as a univariate
polynomial in $y$. Since there exist no $x$-critical points of $\C$
inside $\B$, $C(x)$ is not zero. Hence $g(x,y)$ is degree invariant
over $R$. Also $D(x)=0$ has no roots over $R$. Then, by Theorem
\ref{th-mcc}, $\C$ is delineable over $R$ and $\C_{\B}$ consists of
curve segments starting from certain $L_i$ and ending at certain
$R_j$. Furthermore, these curve segments do not intersect. So
$\sum_{i=1}^l R\#(L_i) = \sum_{j=1}^r R\#(R_i)$. This proves the
lemma.\qed

\begin{figure}[ht]
\centering
\includegraphics[scale=0.30]{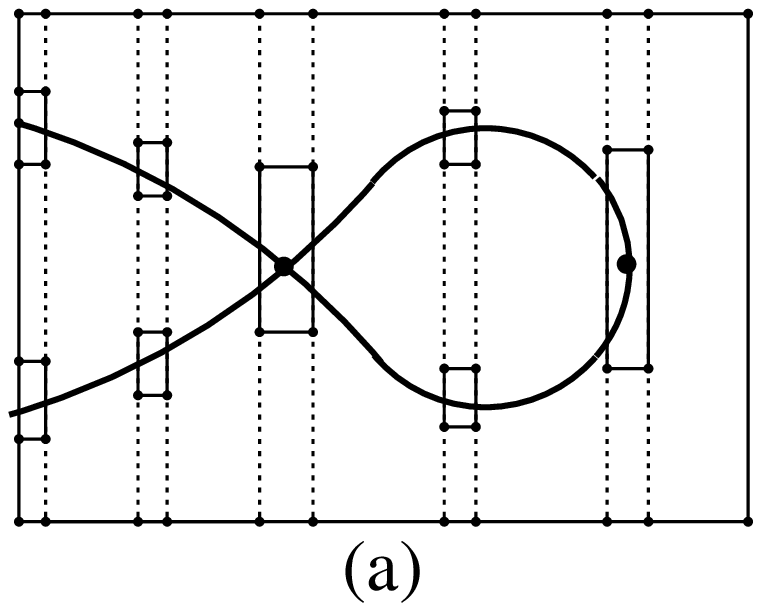}\quad
\includegraphics[scale=0.30]{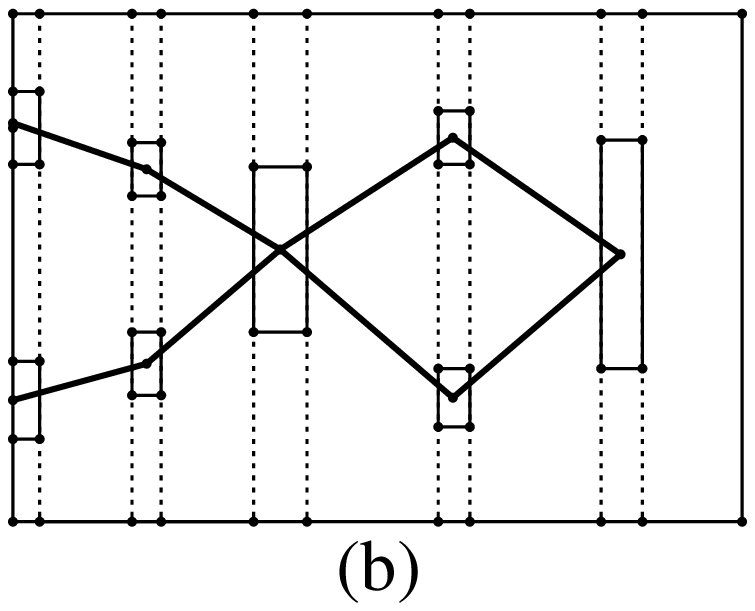}\quad
\includegraphics[scale=0.30]{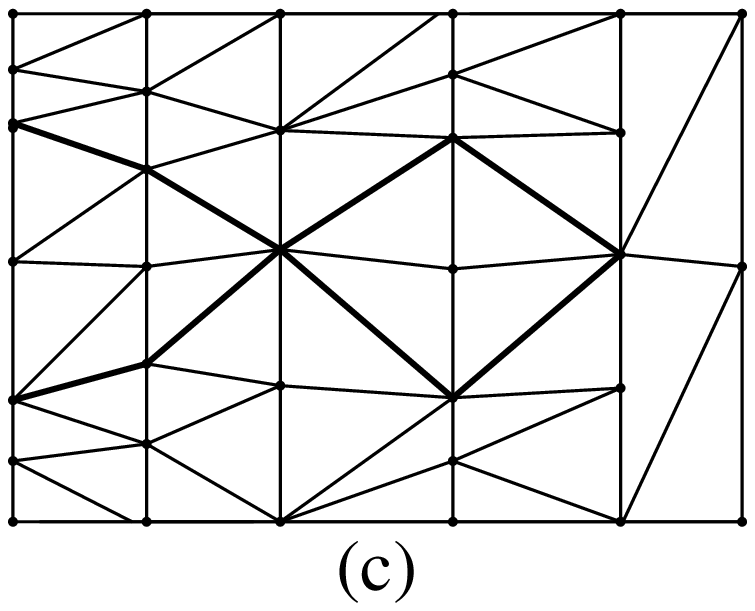}
\caption{Compute topology graph of a curve } \label{fig-egraph-cur}
\end{figure}

A box $\B$ is called a {\bf segregating box} for a point $P$ on $\C$
if $P$ is inside $\B$,  $\B$ is segregating \wrt $\C$, and
$\C_\B\setminus\{P\}$ contains no $x$-critical points of $\C$.
See Fig. \ref{fig-egraph-cur}(a) for an illustration.
It is known that (Theorem 5 in \cite{cad2}):
\begin{lemma}\label{lem-branch}
If $\B=[a, b]\times[c, d]\in\IBQ^2$ is a segregating box of $P$ on
$\C$, then $R\#(P)$ and $L\#(P)$ are the numbers of real roots of
$g(b, y)=0$ and $g(a, y)=0$ in $(c, d)$ respectively. See Fig.
\ref{fig-cur1}(b).
\end{lemma}

The following algorithm computes the branch numbers.
\begin{alg}\label{alg-branch}
{\bf NumCur}$(\PP)$. $\PP$ is a set of points defined by
\bref{eq-ps2}.
Output $R\#(P_{i, j})$ for $0\le i\le s-1$ and $L\#(P_{i, j})$ for
$1\le i\le s$.
\end{alg}
\begin{enumerate}

\item
For $0\le i\le s$, if $g_i(x,y)$ has a factor of the form
$V(x)\in\Q[x]$, let $g_i=g_i/V(x)$.

\item
While $0\in\intbox g_i([a_i,b_i], c_{i, j})$ or $0\in\intbox
g_i([a_i,b_i], d_{i, j})$, repeat $[a_i, b_i]=\ROOTI(h_i(x),[a_i,
b_i]$, $(b_i-a_i)/2).$

\item   Let
$R=\ROOTI(g(b_i, y),[c_{i, j}, d_{i, j}],1)$ and $L=$ $\ROOTI(g(a_i,
y),[c_{i, j}, d_{i, j}],1)$. By Lemma \ref{lem-branch}, $R\#(P_{i,
j})=|R|$ and $L\#(P_{i, j})=|L|$. (See Fig. \ref{fig-cur1}(b))

\end{enumerate}

{\bf Proof of correctness.} Since $\B_{i,j}$ is an isolation box for
$P_{i,j}$, then $g_i(\alpha_i, c_{i, j})g_i(\alpha_i, d_{i,
j})\ne0$.
By Lemma \ref{cor-sl}, the procedure in Step 2 will terminate. At
the end of Step 2, $g_i(x, c_{i, j})g_i(x, d_{i, j})=0$ has no real
roots in $[a_i,  b_i]$, that is, $\B_{i,j}$ is a segregating box for
$P_{i,j}$. The third step is clearly true.\qed

\begin{remark}
In Step 2 of Algorithm \ref{alg-branch}, the boundary points need
special consideration. If the boundary points are on the curve, that
is, if $g(\alpha_i, \YB_1)=0$ or $g(\alpha_i,\YB_2)=0$, we will make
sure that the following condition holds: $\alpha_i$ is the only real
root of $g_i(x,\YB_1)=0$ or $g_i(x,\YB_2)=0$ in $[a_i,b_i]$. Then,
the algorithm also works.
\end{remark}

The following algorithm to compute a topology graph follows the
basic idea in \cite{cad2}. Our main contribution is to use interval
arithmetics instead of algebraic numbers. Also, we do not need
changing the curve to generic positions as done in
\cite{eigenwillig,gon1}
\begin{alg}\label{alg-topcur}
{\bf TopCur}$(g(x,y),\B_2,\epsilon)$. $\C:g(x,y)=0$ is the curve,
$\B_2$ is defined in \bref{eq-b2}, and $\epsilon>0$ is a number.
Output a topology graph $\G=(\PP,\PE)$ which is an {\bf isotopic
meshing} for $\C_{\B_2}$.
 Further, each isolation box $\B$ of a point in
$\PP$ satisfies $|\B|\le \epsilon$.
\end{alg}
\begin{enumerate}

\item Let $\PE = \emptyset$ and
%
$g(x, y) = V(x) g_v(x, y)$,
where $V(x)$ is the factor of $g(x,y)$ in $x$ only.

\item
Let $D(x)=\Res(g_v,\frac{\partial g_v}{\partial y},y)$ be the
resultant of $g_v$ and $\frac{\partial g_v}{\partial y}$.

\item Let $\PP = $\ROOTI$(\Sigma_{21}, \mathbf{B}, \epsilon)\cup
   $\ROOTI$(\Sigma_{22}, \mathbf{B}, \epsilon)$, where
{\small \begin{eqnarray}\label{eqn-hdefinition}
H(x)&=&(x-\XB_1)\cdot(x-\XB_2)\cdot g_v(x, \YB_1)\cdot
 g_v(x, \YB_2)\cdot D(x)\nonumber\\
H_v(x) &=& H(x)/\gcd(H(x),V(x))\\
\Sigma_{21}&=&\{H_v(x),  g_v(x, y)\}\nonumber\\
\Sigma_{22}&=&\{V(x), g_v(x,y)(y-\YB_2)(y-\YB_1)\}.\nonumber
\end{eqnarray}}
%
%
%
%
Assume that $\PP$ is of form \bref{eq-ps2}. See Fig.
\ref{fig-egraph-cur}(a) for an illustration.

\item
Execute Algorithm \ref{alg-branch} to compute $L\#(P_{i, j})\, (1\le
i\le s)$ and $R\#(P_{i, j})\, (0\le i\le s-1)$.

\item
Add an auxiliary line at $x=(b_i+a_{i+1})/2$ and construct the
non-vertical edges. See Fig \ref{fig-cur1}(c) for an illustration.
For $i=0, \ldots,  s-1$, execute the following steps
\begin{enumerate}

\item Let $\QQ_i=$\ROOTI$(\{x-\frac{b_{i}+a_{i+1}}{2}, g_v(x, y)\},
\mathbf{B}, \epsilon)$, where $a_i,b_i$ are from \bref{eq-ps2}. Arrange
the points in $\QQ_i$ bottom up, we have
$\QQ_i=\{Q_{i,1},\ldots,Q_{i,u_i}\}$.
Set $R\#(Q) = L\#(Q) =1$.

\item
Let $\RR_{i}$ be the list of points $P_{i, k}$ arranged bottom up
and point $P_{i, k}$ will be repeated $R\#(P_{i, k})$ times in
$R_{i}$. Similarly,  $\LL_{i}$ is the list of points $P_{i+1, t}$.
By Lemma \ref{prop-22},  $\LL_{i}, \RR_{i}$, and $\QQ_i$ contain the
same number of points. Let $\LL_{i}=(L_1, \ldots, L_{u_i})$,
$\RR_{i}=(R_1, \ldots, R_{u_i})$.

\item For $j=1,\ldots,u_i$, add $(L_j, Q_{i,j}), (Q_{i,j}, R_j)$ to $\PE$.

\item
Let $\PP=\PP\cup\QQ_i$. Still assume that $\PP$ is of form
\bref{eq-ps2}.

\end{enumerate}

\item Add the $x$-vertical edges.
If $\alpha_i$ is a root of $V(x)=0$, add $(P_{i,k},P_{i,k+1}),
k=0,\ldots,s_i$ to $\PE$.

\item Output the topology graph $\G=\{\PP, \PE\}$. The isotopy map
can be constructed in the usual way \cite{pv2}.
\end{enumerate}

\begin{theorem}\label{th-c1}
Algorithm \ref{alg-topcur} computes an isotopic meshing for
$\C_{\mathbf{\B_2}}$.
\end{theorem}
{\em Proof. }
First, we prove that each edge $e\in\PE$ represent exact one curve
segment of $\C$, and for each degree invariant segment of $\C$,
there exist exact one $e\in\PE$ presenting it. Hence $\PE$ and some
$y$-vertical line decompose $\B_2$ into cylindric regions .

It is clear that the curve $\C$ consists of two parts $\C_v:
g_v(x,y)=0$ and  $V(x)=0$. The part $V(x)=0$ consists of straight
lines $x-\gamma_i=0, i=1, \ldots, t$, where $\gamma_i$ are the real
roots of $V(x)=0$. To determine the topology of $\C$, we need only
to find the topology graph $\G_v$ of $\C_v$ and then to add the
lines $x-\gamma_i=0$ to $\G_v$. So we may consider $\C_v$ only.

From Steps 2-4, we know that $\PP$ contains all the $x$-critical
points of the curve $\C_v$ and the {\em boundary points} which are
the intersection points of $\C$ and the boundaries of $\B_2$.
In Steps 6 and 7, we add auxiliary points $\QQ_i$ to $\PP$. Since
points in $\QQ_i$ are not critical points of $\C$, we have $R\#(Q) =
L\#(Q) =1$ for $Q\in \QQ_i$. This makes sure that all the edges
$(L_j,Q_{i,j})$ and $(Q_{i,j},R_j)$ are distinct.

Let $B_i = (\alpha_i,\alpha_{i+1})\times[\YB_1,
\YB_2],i=0,\ldots,s-1$. We need only to show that $\C_v$ and $\G_v$
have the same topology in $B_i$.
Let $S_i$ be the interval $(\alpha_i,\alpha_{i+1})$. Then $D(x)$
does not vanish on any point of $S_i$. As a consequence, $g_v(x,y)$
must be degree invariant on $S_i$. By Theorem \ref{th-mcc},
$g_v(x,y)$ is delineable over $S_i$ and $\G_v$ is obtained by
replacing a curve segment of $\C_v$ in $B_i$ by a line segment with
same end points. It is clear that $\C_v$ and $\G_v$ have the same
topology. We are going to make  explicit the isotopy from $\PE$ to
$\C_{\mathbf{\B_2}}$. Let $\G=(\PP,\PE)$ be a topology graph for
curve $\C_{\mathbf{\B_2}}$, and $\PP$ of form \bref{eq-ps2}.
Let $P_{i,j}=(\alpha_{i},\beta_{i,j})$ be of form \bref{eq-ps2}. Let
$Q_{i,j}=(\tau_i, \rho_{i,j})$ where $\tau_i=\frac{a_i+b_i}{2},
\rho_{i,j}=\frac{c_{i,j}+d_{i,j}}{2}$.
Then $\G$ decomposes $\B_2$ into cylindrical regions $\cup_{i,j}
R_{i,k}$, where $R_{i,j}$ is bounded by $[\tau_i,\tau_{i+1}]$ in the
$x$-direction  and by $f_{1}=(Q_{i,u},Q_{i+1,v})$ and
$f_{2}=(Q_{i,s},Q_{i+1,t})$ for ceratin $u,v,s,t$. Note that
$R_{i,k}$ could be a triangle or a quadrilateral.

First,  we consider one cylindrical region $R_{i,k}$ defined as
above. Let $e_1 =(P_{i,u},P_{i+1,v}))$ and $e_2
=(P_{i,s},P_{i+1,t})$.
Without loss of generality, assume
$\alpha_1<\alpha_2,\beta_{1,1}<\beta_{1,2}$.
According to the correctness prove of Algorithm \ref{alg-topcur},
$g_v(x,y)$ is delineable over $[\alpha_1,\alpha_2]$, we can find two
root functions $\theta_i(x)$ of $g_v$ on $[\alpha_1,\alpha_2]$
corresponding to the two curve segments $C(e_1)$ and $C(e_2)$.
Denote $y=\delta_i(x), x\in [\alpha_1,\alpha_2]$ to be the
definition functions of line segments $e_i$ and $y=\varphi_i(x),
x\in [\tau_1, \tau_2]$ to be the definition functions of line
segments $f_i$. Consider the maps:
$$F_1: ([\alpha_1,\alpha_2]\times\R)\times [0,1]\rightarrow \R^2$$
 defined by
$$\small \begin{array}{rl}
&(x,\lambda\delta_1(x)+(1-\lambda\delta_2(x)),t)\\
\rightarrow&(x,\lambda(t\theta_1(x)+(1-t)\delta_1(x))+(1-\lambda)(t\theta_2(x)+(1-t)\delta_2(x)))
\end{array}$$
and
$$F_2: ([\tau_1,\tau_2]\times\R)\times[0,1]\rightarrow [\alpha_1,\alpha_2]\times\R$$
defined by
$$\small \begin{array}{rl}
&(x,\lambda\varphi_1(x)+(1-\lambda\varphi_2(x)),t)\\
\rightarrow&(x',\lambda(t\delta_1(x')+(1-t)\varphi_1(x))+(1-\lambda)(t\delta_2(x')+(1-t)\varphi_2(x))),\end{array}$$
where
$x'=\alpha_1+\frac{x-\tau_1}{\tau_2-\tau_1}(\alpha_2-\alpha_1)$.

The map $F_2$ is a homeomorphism from $[\tau_1,\tau_2]\times\R$ to
$[\alpha_1,\alpha_2]\times\R$ and $F_1$ is a homeomorphism from
$[\alpha_1,\alpha_2]\times\R$ to itself. So the composed map
$F_{i,k}:=F_1\circ F_2$ is a homeomorphism from
$[\tau_1,\tau_2]\times\R$ to $[\alpha_1,\alpha_2]\times\R$ and it
deforms $f_i$ to $C(e_i)$ continuously.
Extend this map to $\R^2\times[0,1]$ by setting it to be the
identity map outside $R_{i,k}$, we obtain an isotopy from line
segments $f_1\cup f_2$ to the curve segments $C(e_1)\cup C(e_2)$.

Now we consider the whole topology graph $\G$. For each cylindrical
region $R_{i,k}$, we can construct an isotopy $F_{i,k}$ as above.
Consider the following map:
$$F: \R^2\times[0,1]\rightarrow\R^2$$
denoted by
\begin{equation}\label{eq-iso}
F(P,t)=\left\{\begin{array}{ll} F_{i,j}(P,t), &P\in R_{i,j},\\
id, &P\in \R^2\setminus \B_2
\end{array} \right.
\end{equation}
Note that $F_{i,j}|_{R_{i,j}\cap R_{u,v}}=F_{u,v}|_{R_{i,j}\cap
R_{u,v}}$, and $F_{i,j}|_{R_{i,j}\cap (\R^2\setminus\B_2)}=id$ for
all $i,j,u,v$. $(\G , F)$ is an isotopy for
$\C_{\mathbf{\B_2}}$.\qed

As a consequence of the above proof, we have
\begin{corollary}\label{cor-21}
Let $G=(\PP,\PE)$ be a topology graph of the curve $\C_{\B_2}$
obtained by Algorithm \ref{alg-topcur}. Then all the singular points
of $\C_{\B_2}$ are in $\PP$ and $g(x,y)$ is $y$-degree invariant
over the intervals $(\alpha_i,\alpha_{i+1}),i=0,\ldots,s-1$.
\end{corollary}


When computing the topology of a surface, we need to introduce the
concept of extended topology graph.
An {\bf  extended topology graph} associate with a box
$\mathbf{B}_2$ is a triplet $\EG= \{\EP,\EE,\EC\}$ where
$\{\EP,\EE\}$ is a topology graph and
$\EC=\{(P_1,P_2,P_3) \,|\,
P_i\in \EP,$ $(P_1,P_2),$ $(P_2,P_3)$, $(P_3,P_1)$ $\in \EE \}$ is a
set triangular cells in $\B_2$.
We further assume that  the cells in $\EC$ are disjoint except on
their edges and  provide a cover for $\B_2$.
%

We can obtain an extended topology graph of a curve from a topology
graph by adding more auxiliary points and edges.

\begin{alg}\label{alg-etopcur} {\bf ETopCur}$(g(x,y),\B_2,\rho)$
The input is the same as Algorithm \ref{alg-topcur}. The output is
an extended topology graph of $\C_{\B_2}$. (See Fig.
\ref{fig-egraph-cur}(c) for an illustration)
\end{alg}
\begin{enumerate}
\item 
 Let $\G=\{\PP,\PE\}={\bf TopCur}(g(x,y),\B_2,\rho)$.

\item Let $\EP=\PP$. For $i=0,\ldots,s$, add
$(\alpha_i,\YB_1)$ and $P_{i,s_i}=(\alpha_i,\YB_2)$ to $\EP$ if they
are not in it.

\item For points $P_{i,j},j=0,\ldots,s_i$, let
$[a_i,b_i]\times[c_{i,j},d_{i,j}]$ be the isolation box for
$P_{i,j}$. Add $N_{i,j}=(\alpha_i, (d_{i,j}+c_{i,j+1})/2),
j=0,\ldots,s_i-1$ to $\EP$. We still assume that $\EP$ is of form
\bref{eq-ps2}.

\item Let $\EE=\PE$. For $j=0,\ldots, s_0-1$, add the edges
$(P_{0,j},N_{0,j}),(N_{0,j},P_{0,j+1})$ to $\EE$.

\item For each $0\le i\le s-1$, add the edges
$(P_{i,0},P_{i+1,0})$ and $(P_{i,s_{i}},P_{i+1,s_{i+1}})$ to $\EE$.
Then the edges in $\EE$ divide  the rectangular region
$B=[\alpha_i,\alpha_{i+1}]\times[\YB_1,\YB_2]$ into triangular and
quadrilateral regions. We will subdivide these regions into
triangular regions such that each point in $\EP$ is the vertex of at
least one triangles.

\item
Let $\EC=\emptyset$. For any two adjacent edges $e_1=(P_{1,1}$,
$P_{2,1}), e_2=(P_{1,2},P_{2,2})$ inside $B$, execute Steps 7 and 8.
%

\item If $P_{1,1}\neq P_{1,2}$, there exists one point $N_1$ added
before in Step 3 between $P_{1,1}$ and $P_{1,2}$. Furthermore,
\begin{itemize}
\item If $P_{2,1}\neq P_{2,2}$, there exists one point $N_2$ between $P_{2,1}$ and
$P_{2,2}$. $P_{1,1}, P_{1,2}, P_{2,2}, P_{2,1}$ form a quadrilateral
region. We can divide the quadrilateral region into four triangles.
Add the edges $(P_{2,1},N_2),(N_2, P_{2,2}), (P_{1,2},N_2),
(N_1,N_2), (N_1,P_{2,1})$ to $\EE$. Add the triangles $(P_{1,1},N_1,
P_{2,1})$, $(N_1,$ $P_{2,1},N_2), (N_1,P_{1,2},N_2)$,
$(P_{1,2},N_2,P_{2,2})$ to $\EC$.

\item If $P_{2,1} = P_{2,2} = P$, $P_{1,1}, P_{1,2}, P$ form a
triangular region. We can divide the triangular region into two
triangles. Add the edges $(P_{1,1},N_1), (N_1,P_{1,2}), (N_1,P)$ to
$\EE$. Add the triangles $(P_{1,1},N_1,P), (N_1,P_{1,2},P)$ to
$\EC$.
\end{itemize}

\item If $P_{1,1} = P_{1,2} = P$, then there must exist a point $N_2$
added before in Step 3 between $P_{2,1}$ and $P_{2,2}$. $P, P_{2,2},
P_{2,1}$ form a triangular region. We can divide the quadrilateral
region into two triangles. Add the edges $(P_{2,1},N_2),(N_2,
P_{2,2}), (P,N_2)$ to $\EE$. Add the triangles $(P,P_{2,1},N_2),
(P,P_{2,2},N_2)$ to $\EC$.

\item Output $\EG=\{\EP,\EE,\EC\}$.
\end{enumerate}

{\bf Remark.} The purpose to add points $N_{i,j}$ in Step 3 is to
make sure that topology representation for surfaces possible.
These points has similar function as the the auxiliary points added
in Step 6 of Algorithm \ref{alg-topcur}. Hence, they are also called
{\bf auxiliary points}.
Figure \ref{fig-egraph-cur} is an extend topology graph of the curve
$G(x,y)=x\cdot y\cdot(16 x^2+ 16 y^2 -49)=0$.

Let $\EG=\{\EP, \EE, \EC\}$ be an extend topology graph of
$\C_{\mathbf{B}_2}$ and $e=(P_1, P_2)\in\EE$.
If $e$ corresponds to a curve segment of $\C_{\B_2}$,  we use $C(e)$
to represent the corresponding curve segment; otherwise, we use
$C(e)$ to represent the line segment $P_1P_2$. Let
$I(e)=C(e)\setminus\{P_1,P_2\}$.
For a $c\in\EC$, we use $R(c)$ and $I(c)$ to denote the cell and
interior of the cell represented by $c$ respectively.

\subsection{Compute $\epsilon$-meshing for plane curve}
\label{sec-c2}

The meshing given in Section \ref{sec-c1} has no guarantee of
precision. In this section, we will show how to compute a meshing
for a curve to any given precision.

Let $\G=(\PP,\PE)$ be a topology graph for a curve $\C$ inside a box
$\B_2$ defined in \bref{eq-b2}. Assume that $\PP$ is of form
\bref{eq-ps2}. Consider the two disjoint regions $\SB_2$ and $\NB_2$
of $\B_2$:
\begin{eqnarray}
\SB_2&=&\cup_i \SB_2^i, \NB_2=\cup_j \NB_2^j, \hbox{ where }\nonumber\\
\SB_2^i &=& (a_i,b_{i})\times[\YB_1,\YB_2],i=0,\ldots,s\label{eq-i1}\\
\NB_2^j &=&[b_j,a_{j+1}]\times[\YB_1,\YB_2],j=0,\ldots,s-1.\nonumber
\end{eqnarray}
Then, $\C_{\B_2}\subset \SB_2\cup\NB_2$ and is smooth in $\NB_2$.
%

The idea of our algorithm is to determine the topology of the curve
in the region $\SB_2$ with Algorithm \ref{alg-topcur}, to determine
the topology of the curve in the region $\NB_2$ with a modified
marching cube method of Pantinga-Vegter \cite{pv1}, and to compute
the adjacency information on the border lines $x=a_i, x=b_i$. We
could use the marching cube method in $\NB_2$ because $\C$ has no
singular point in it.

\vskip-6pt
\begin{figure}[ht]
\begin{minipage}{0.57\textwidth}
\begin{center}
\includegraphics[scale=0.3]{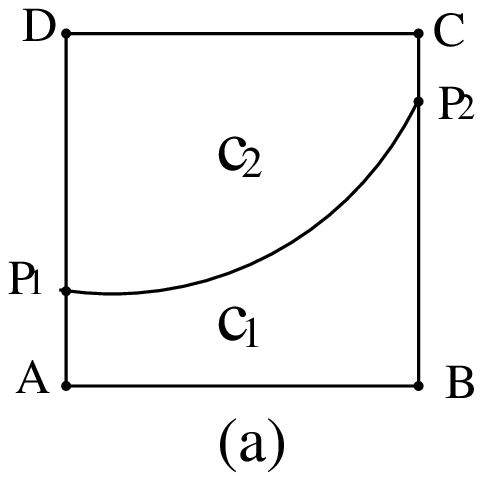}
\includegraphics[scale=0.3]{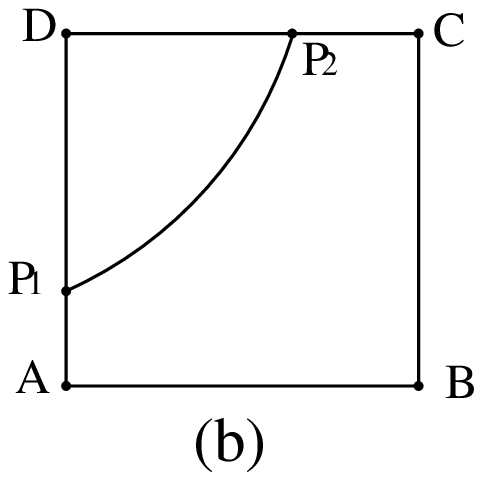}\quad
\includegraphics[scale=0.26]{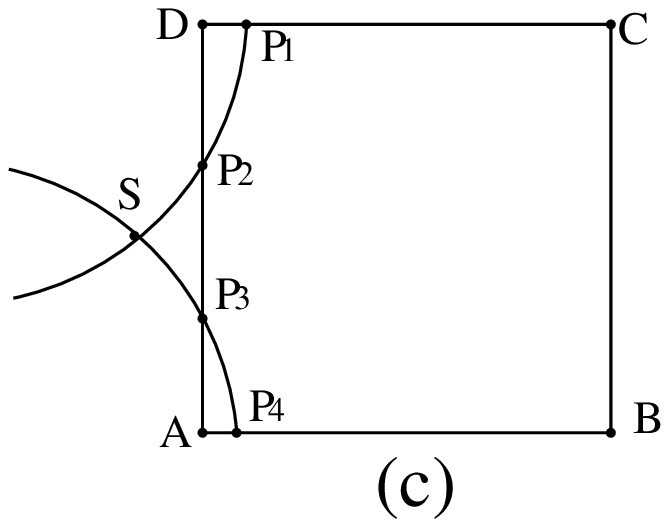}
\includegraphics[scale=0.27]{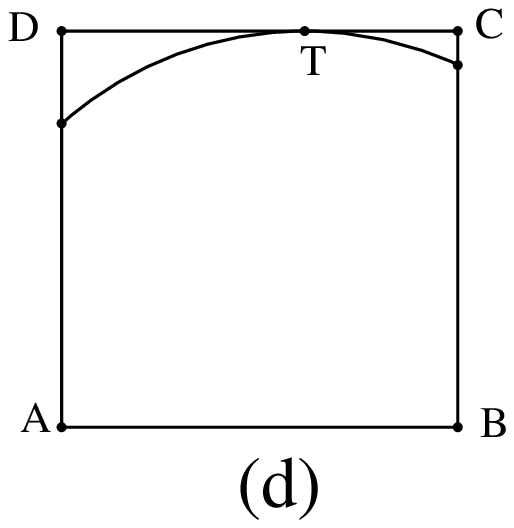}
\caption{Nice boxes: (a), (b). Boxes in (c), (d)  not
nice.}\label{fig-ms-c}
\end{center}
\end{minipage}
\begin{minipage}{0.385\textwidth}
\centering
\includegraphics[scale=0.28]{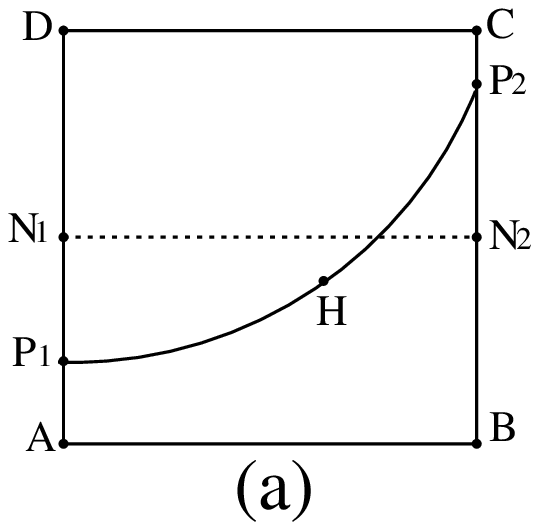}\hskip14pt
\includegraphics[scale=0.34]{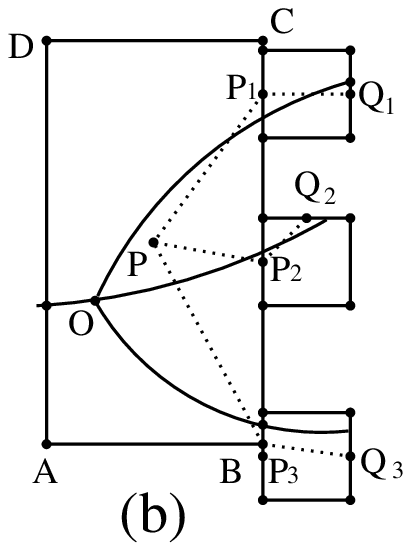}
\caption{Meshing  curve segments} \label{fig-ms-m}
\end{minipage}
\end{figure}

In order for the above idea to work, we need to modify the
Pantinga-Vegter method such that each output box  contains only one
curve segment of $\C$, as shown in Fig. \ref{fig-ms-c}(a) and (b).
Such boxes are called {\bf nice boxes}.

The original Pantinga-Vegter method could output a box containing
two curve segments and this will cause problems when the box is near
a singular point, as shown in Fig. \ref{fig-ms-c}(c).
A point is called a {\bf $y$-extremal point} of curve $\C$ if $\C$
achieves a local extremum value at this point in the $y$-direction.
Pantinga-Vegter's method could output a box shown in Fig.
\ref{fig-ms-c}(d).

To make the process precise, we introduce the following definition.
An  {\bf $\epsilon$-meshing graph} of a curve $\C$ is a triplet
$\MM=\{\PP,\PE,\MB\}$ where
$(\PP,\PE)$ is a graph whose vertices are with rational numbers as
coordinates and whose {\bf edges are the  meshes} for $\C$;
$\MB$ is a set of nice boxes and segregating boxes of singular
points of $\C$ such that for each $e\in\PE$, there exists a
$\B_e\in\MB$ with the property: $|\B_e|<\epsilon$ and $\C\cap\B_e$
is a connected curve segment of $\C$ (See Fig. \ref{fig-ms-m}). 
In Fig. \ref{fig-ms-m}(a), $\PP=\{N_1,N_2\}, e=(N_1,N_2), \B_e=ABCD$
forms a meshing graph for curve segment $C(e) =P_1HP_2$.

It is easy to show that an $\epsilon$-meshing graph for a curve $\C$
provides an {\bf $\epsilon$-meshing} for $\C$ according to the
definition given in Section 2.

\begin{alg}\label{alg-mpv2}{\bf MPV2}$(g(x,y),\B_2,\epsilon).$
Input: $\C: g(x,y)=0$ is a curve with no $x$-critical points and no
$y$-extremal points in box $\B_2$.
Output  an  $\epsilon$-meshing graph $\G=\{\PP,\PE,\MB\}$ for
$\C_{\B_2}$.
\end{alg}
We need only add some extra criterions for the boxes:
(1) For each edge $(A,B)$ of $\B$, if $0\in \intbox g((A,B))$ and
$g(A)g(B)>0$, we continue to subdivide $\B$.
%
(2) For each box $\B$, if $|\B|>\epsilon$, we continue to subdivide
$\B$.

Since $\C$  has no $x$-critical points and no $y$-extremal points in
box $\B_2$, a box like the one in Fig. \ref{fig-ms-c}(d) does not
exist and the algorithm will terminate.

\begin{figure}[ht]
\begin{center}
\includegraphics[scale=0.3]{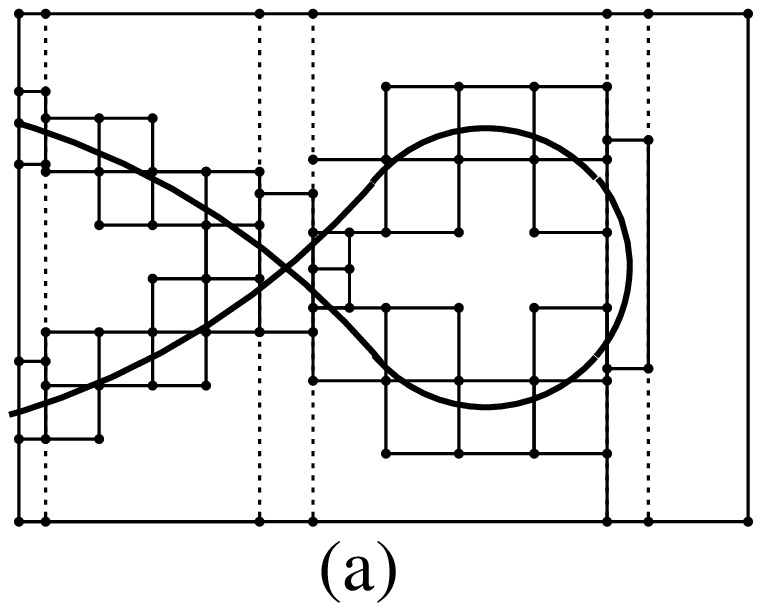}\quad
\includegraphics[scale=0.3]{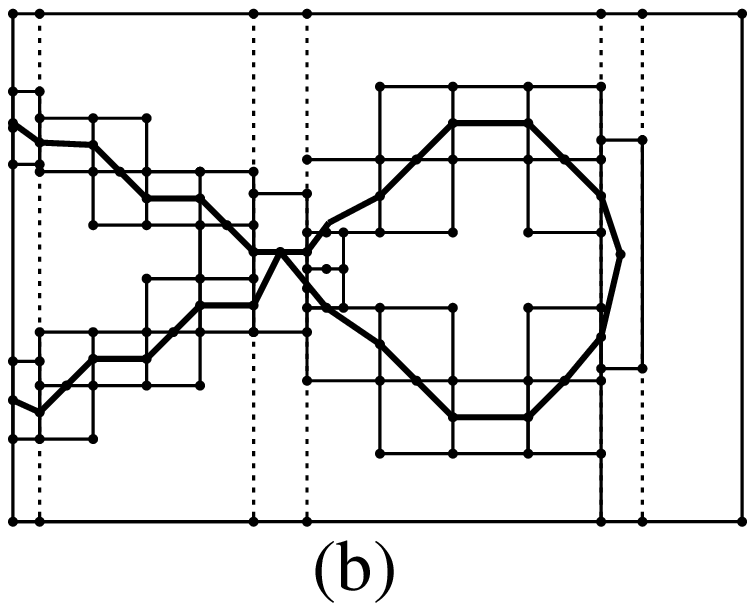}
\includegraphics[scale=0.30]{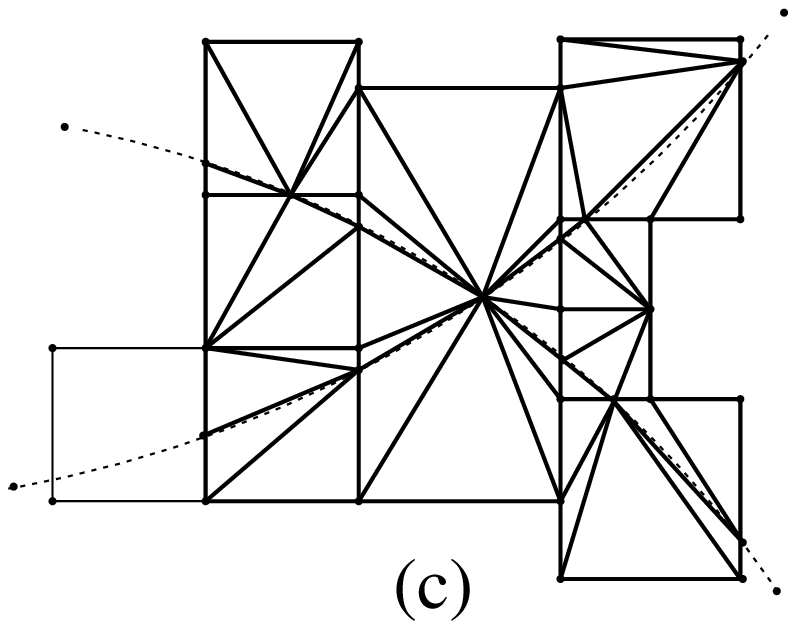}
\caption{$\epsilon$-meshing for a curve} \label{fig-ms-ce}
\end{center}
\end{figure}

Now, we can give the meshing algorithm for curves.
\begin{alg}\label{alg-atopcur}
{\bf ATopCur}$(g(x,y),\B_2,\epsilon)$. The input is the same as that
of Algorithm \ref{alg-topcur}. Output an $\epsilon$-meshing graph
for $g(x,y)=0$.
\end{alg}
\begin{enumerate}
\item
Execute the first four steps of Algorithm \ref{alg-topcur} with
input ($g(x,y),\B_2,\epsilon)$.
We need to modify Algorithm \ref{alg-topcur} as follows:
Let $g_u(x,y) = g_v(x,y)/U(y)$ where $U(y)$ is the gcd of the
coefficients of $g_v(x,y)$ as a univariate polynomial in $x$; and
use $H(x)\cdot \Res(g_u,\frac{\partial g_u}{\partial x},y)$ as the
new $H(x)$ in \bref{eqn-hdefinition}.

We need $V(x), g_v(x,y),$ and $\G_1=\{\PP_1,\MB_1\}$ from Algorithm
\ref{alg-topcur}, where $\MB_1$ is the segregating boxes for the
points in $\PP_1$.

\item Compute $\G_2=\{\PP_2,\PE_2,\MB_2\}$={\bf
MPV2}($g(x,y),\NB_2, \epsilon$). The modification in Step 1 makes
sure that $\C$ has no $x$-critical points and $y$-extremal points in
$\NB_2$.

\item
 Compute the connection between the boxes computed by Step S1 and Step S2. (Fig. \ref{fig-ms-m}(b) shows how to mesh $\C$ near a singular point
$O$ with $\B=ABCD$ as its segregating box.  Fig. \ref{fig-ms-ce}(b) provides a global picture for meshing a curve.)
 For $i=0,\ldots,s$, consider the adjacency information on the border
lines $x=a_i, b_i$. We only consider $x=b_i$.
For each $P\in\PP$ and its segregating box $\B=[a,b]\times[c,d]\in
\MB$, do the following
\begin{enumerate}

\item
Let $\EB_k=[b,c_k]\times[e_k,f_k]\in\MB_2$ be the boxes satisfying
$\B\cap\EB_k\ne\emptyset$ and $g(b,\hat{e}_k)g(b,\hat{f}_k)<0$,
where $\hat{e}_k = \min\{e_k, c\}$ and $\hat{f}_k = \max\{f_k, d\}$.
As a consequence, $\C$ passes through these $\EB_k$ through the
interval $[b_i,b_i]\times[e_k,f_k]$.

\item Let  $Q =((a+b)/2,(c+d)/2)$, $m_k = (\hat{e}_k+\hat{f}_k)/2$.

\item
Add the edge $e=(Q,(b,m_k))$ to $\MM$. Add $\B_e = \B$ to $\MM$.
\end{enumerate}

\item Add the meshes for the straight lines defined by
$V(x)=0$.
\item
Output the meshing graph $\MM$.
\end{enumerate}

\vskip-6pt

\begin{theorem}\label{th-m1}
Algorithm \ref{alg-atopcur} terminates and computes an
$\epsilon$-ambient meshing for $\C_{\mathbf{\B_2}}$.
\end{theorem}

{\em Proof. }
By Lemma \ref{th-c1}, $\MM\cap \SB_2$ is the isotopic meshing of
$\C\cap\SB_2$. Marching cube method compute the isotopic meshing of
$\C\cap\NB_2$. Hence, $\MM$ is a isotopic meshing for $\C_{\B_2}$.
Furthermore, for each $e\in\PE\cap \NB_2$, $C(e)\subseteq \B_e,
|\B_e|<\epsilon$ and each part  of $\C_{\B_2}$ around the singular
point $P$ is contained in the segregating boxes $\B_P,
|\B_P|<\epsilon$ of $P$. Therefore, for any point $P\in \MM$,
$F(P,1)$ and $P$ are in the same box $\B_{e}$ with
$|\B_{e}|<\epsilon$, so $\parallel F(P,1)-P\parallel <\epsilon$.
This gives a proof of Theorem \ref{th-m1}. \qed


In order to compute the $\epsilon$-meshing for surfaces, we need to
add more information to the $\epsilon$-meshing graph.
Let $\MM=\{\PP,\PE,\MB\}$ be a meshing graph for a curve $\C$. Then
an {\bf extended meshing graph} $\PE\M=\{\EP,\EE,\EC\}$ for $\C$ in
$\MB$ can be defined similarly as the extended topology graph. The
difference is that $\EC$ provides a triangular decomposition for
$\MB$.
%
%

The following algorithm computes an extended meshing graph.
\begin{alg}\label{alg-emtopcur}
{\bf METopCur}$(g(x,y),\G_1,\G_2)$. $\G_1=\{\PP_1,\MB_1\}$ where
$\PP_1$ is the set of  points on $\C:g(x,y)=0$ of form \bref{eq-ps2}
and $\MB_1$ is the set of their segregating boxes.
$\G_2=\{\PP_2,\PE_2,\MB_2 \}$ is an $\epsilon$-meshing of the curve
$\C$ in $\NB_2$ defined in \bref{eq-i1}. Output an extended meshing
graph for $\C$ in $\MB_1\cup\MB_2$. (Fig. \ref{fig-ms-ce}(c) is the
extended meshing graph for the box in the center of Fig.
\ref{fig-ms-ce}(a) and its surrounding boxes.)
\end{alg}
\begin{description}
\item[S1] Let
$\EP=\emptyset,\EE=\emptyset,\EC=\emptyset$.

\item[S2] For any $\B=[a,b]\times[c,d]\in \MB_1$, it is the segregating box of  one point $P\in
\PP_1$. Compute the extended topology in $\MB_1$.

\begin{enumerate}
\item   Compute
\begin{eqnarray*}
 \{Q_1,\ldots, Q_l\}&=&\mbox{\bf RootIsol}(\{x-a,g(a,y)\}, [c,d], 1)\\
 \{T_1,\ldots,T_r\}&=&\mbox{\bf RootIsol}(\{x-b,g(b,y)\}, [c,d], 1).
\end{eqnarray*}
Denote $A_i,i=1,\ldots,4$ to be the four vertices of $\B$. Denote
$B_j,j=1,\ldots,p$  to be the  points on the edge of $\B$ which are
the vertices of boxes adjacent to $\B$.(Note that if $l>1$($r>1$),
there exists some point $B_k$ between $Q_i$ and $Q_{i+1}$($T_i$ and
$T_{i+1}$)).

\item $\TT\PP=\emptyset$. Add points $Q_i,T_j,A_k,B_p$ into $\TT\PP$.

\item Denote the sets of points $L=\{L_1,\ldots,s\}$ and
$R=\{R_1,\ldots,R_t\}$ where $L$ is the  points in $\TT\PP$ which
are on the left edge of $\B$ sorted from bottom to up and $R$ the
points in $\TT\PP$ which are on on the right edge of $\B$ sorted
from bottom to up.

\item Add point $P$ and all points in $\TT\PP$ into $\EP$. Add edges
$(P,L_i),(P,R_j)$ into $\EE$. Add triangular cell
$(P,L_i,L_{i+1}),(P,R_i,R_{i+1})$ and $(P,L_1,R_1),(P,L_s,R_t)$ and
$(L_i,L_{i+1})$, $(R_j,R_{j+1})$ to $\EC$.
\end{enumerate}

\item[S3] For any boxes  $\B\in\MB_2$. Compute the extended topology in $\B$.
For any line segment $e\in \PE_2$ with $\B=[a,b]\times[c,d]\in\MB_2$
containing it.
 Doing the following operations(There are six
conditions that $e$ divide $\B$ into two parts, see fig
\ref{fig-ms-c}. We can distinguish them according to $\PP_2$ and
$\MB_2$. Here we consider the condition (a), the other conditions
are dealt with in the similar way).

\begin{enumerate}
\item Compute $Q=\mbox{\bf RootIsol}(\{x-a,g(x,y)\},\B,\epsilon/4)$ and
$T=\mbox{\bf RootIsol}(\{x-b,g(x,y)\},\B$, $\epsilon/4)$. Obviously,
$Q$ and $T$ both contain only one point. We still call them $Q$ and
$T$. Denote $A_i,i=1,\ldots,4$ to be the four vertices of $\B$.
Denote $B_j,j=1,\ldots,p$ to be  points on one edge of $\B$ which
are the vertices of boxes adjacent to $\B$.

\item $\TT\PP=\emptyset$. Add points $Q,T,A_i,B_j$ into $\TT\PP$.

\item Add all points in $\TT\PP$ into $\EP$. Similar to the forth
step in Step {\bf S2}, we can decompose $\B$ into triangular cells
and insert these cells into $\EC$, and insert corresponding edges
into $\EE$ such that each point in $\TT\PP$ connects to at least
another point in this set.

\end{enumerate}

\item[S4] Output  $\PE\M=\{\EP,\EE,\EC\}$.
\end{description}

\section{Topology of  surface}
In this section, an algorithm will be given to compute a polyhedron
with triangular faces, which is isotopic to a given surface.

\subsection{Outline of the algorithm}
We use a polyhedron with triangular faces to represent the topology
of a surface.
A {\bf topology polyhedron} is a triplet $\P= \{\SP, \SE, \SF\}$
where $\SP$,  $\SE$,  and  $\SF$ are defined below.
\begin{itemize}
\item
$\SP$ is a set of 3D points determined by a triangular system
$\Sigma_{i}$ and an isolation boxes $\B_{i,j,k}$:
{\small
\begin{eqnarray}\label{eq-ps3}
 \SP &=& \{P_{i, j, k},  0\le i\le s,  0\le j \le
s_i,  0\le k \le t_{i, j}\}\nonumber\\
\Sigma_{i} &=& \{h_{i}(x), g_{i}(x, y), f_i(x, y, z)\}\\
\B_{i,j,k}&=&[a_i, b_i]\times[c_{i, j}, d_{i, j}]\times[e_{i, j, k},
f_{i, j, k}]\in\IBQ^3.\nonumber
\end{eqnarray}}
where $P_{i, j, k}=(\alpha_i,  \beta_{i, j}, \gamma_{i, j, k})$
satisfy $\alpha_0<\cdots<\alpha_s$, $\beta_{i, 0}<\cdots<\beta_{i,
s_i}$, and $\gamma_{i, j, 0}<\cdots<\gamma_{i, j, t_{i, j}}$.
Point $P_{i, j, k}$ is said to be {\bf lifted } from the plane point
$P_{i, j} = (\alpha_i,  \beta_{i, j})$.
$P_{i, j}$ is said to be the {\bf projection } of $P_{i,j,k}$.

\item $\SE=\{(P_1, P_2) | P_1, P_2\in\SP, $ such that either
$P_1=P_{i, u, v}, P_2=P_{i+1, p, q}$ \hbox{ or } $P_1=P_{i, u, v, },
P_2=P_{i, u+1, t}\}$.
We further assume that any two edges do not intersect except at the
end points.

\item $\SF=\{(P_1, P_2, P_3) | P_1, P_2, P_3\in\SP \}$  such that
its three edges are in $\SE$.
We further assume that any two faces do not intersect except at the
edges.

\end{itemize}

Let $\S: f(x, y, z)=0$ be an algebraic surface, where $f(x, y, z)\in
\Q[x, y, z]$ is square free.
A point $P_0$ is a {\bf critical point} of $\S$ if
$f(P_0)=f_z(P_0)=0$. Write $f$ as a univariate polynomial in $z$:
$f(x, y, z)=f_d(x,y)z^d + \cdots + f_0(x,y).$ $f_d(x,y)$ is called
the {\bf leading coefficient} of $f(x,y,z)$.
We further assume that
\begin{equation}\label{eq-co1}
f_d(x,y)=\cdots= f_0(x,y)=0 \hbox{ have no common
zeros.}\end{equation}
Geometrically, this means that $\S$ does not contain a line parallel
to the $z$-axis. We will consider surfaces that do not satisfy this
condition in Section \ref{sec-gen}.

Similar to the case of algebraic curves, we will consider the
topology of $\S$ in a bounding box
\begin{equation}\label{eq-b3}\B_3=[\XB_1, \XB_2]\times[\YB_1, \YB_2]\times[\ZB_1,
\ZB_2]\in\IBQ^3.\end{equation}

Let {\small
\begin{eqnarray}
D(x, y)&=&\res(f,\frac{\partial f}{\partial z},z)\label{eq-D} \\
G(x,y)
&=& \hbox{sqrfree}( D(x, y) f(x, y, \mathcal{Z}_1)f(x, y,
\mathcal{Z}_2))\label{eq-projcur}
\end{eqnarray}}
where sqrfree$(P(x,y))$ is the square free part of $P(x,y)$.
The plane curve $G(x, y)=0$ is called  the {\bf projection curve} of
$\S$.

To determine the topology of a surface is to find a topology
polyhedron with the same topology as the surface. We first give an
outline of the algorithm, which consists of four main steps.
\begin{description}
\item[S1]
Compute an extended topology graph $\EG= \{\EP,\EE,\EC\}$ of the
projection curve of $\S$ in $\B_2$ defined in \bref{eq-b2}.

\item[S2]
{\bf Determine $\SP$.} For any $P\in\EP$, determine the intersection
points of $\S$ and the line segment $P\times [\ZB_1, \ZB_2]$.

\item[S3]
{\bf Determine $\SE$.}
For each edge $e\in\EE$, compute the intersection  of $\S$ and the
cylindrical surface patch $I(e)\times [\ZB_1, \ZB_2]$, which are
delineable curve segments of $\S$ whose end points are in $\SP$. We
will use line segments in $\SE$ to represent these curve segments.
See Fig. \ref{fig-top-spcur}.

\item[S4]
{\bf Determine $\SF$.} For each $c\in \EC$, compute the intersection
of $\S$ and the prism $I(c)\times[\ZB_1, \ZB_2]$, which are
delineable surface patches of $\S$ whose edges are in $\SE$. We will
use triangular faces in $\SF$ to represent these surface patches.
See Fig. \ref{fig-top-surfp}.

\end{description}

\subsection{Theoretical preparations for the algorithm}
\label{sec-s2}
In the outline of the algorithm given in the preceding section, Step
{\bf S}1 has been solved in Section \ref{sec-c1}. Step {\bf S}2 can
be solved with Algorithm \ROOTI. We will explain Steps
 {\bf S}3 and {\bf S}4 below.

Roughly speaking, Step {\bf S3} is to determine the topology of the
spatial curve defined by $f(x,y,z)=G(x,y)=0.$
The following result,  which is a consequence of Theorem
\ref{th-mcc}, allows us to determine the singularities of this curve
easily.
\begin{lemma}\label{lm-3c}
Use the notations introduced above. For each edge
$e=(P_1,P_2)\in\EE$, $f(x,y,z)=0$ is delineable over
$I(e)=C(e)\setminus\{P_1,P_2\}$.
\end{lemma}
{\em Proof.} Let $\C:G(x,y)=0$ be the projection curve of $\S$ and
$D(x,y)$ the discriminant of $f$ \wrt  $z$. Since $I(e)$ is a
continous curve segment of $\C$, $G$ is order-invariant on $I(e)$.
From \bref{eq-D},  $D(x,y)$ is order-invariant on $I(e)$.
Since condition (\ref{eq-co1}) holds, $f$ does not vanish
identically on any point of $xy$-plane. So, $f$ does not vanish
identically on $I(e)$.
Now, we will prove that $f$ is degree-invariant on $I(e)$.
It is clear that all the singular points of $\C$ are in $\EP$. Then
$f_d(x,y)$ is either identically zero on $I(e)$ or does not vanish
on any point on $I(e)$. So we can conclude that $f_d(x,y)$ is
sign-invariant on $I(e)$. By Theorem \ref{th-brown}, $f$ is
degree-invariant on $I(e)$ . By Theorem \ref{th-mcc}, $f$ is
delineable on $I(e)$. \qed

As a corollary, we have
\begin{corollary}\label{cor-3c1}
For $e=(P_1,P_2)\in\EE$, the intersection of $\S$ and $I(e)\times
[\ZB_1, \ZB_2]$ consists of disjoint curve segments of $\S$ whose
end points are in $\SP$.
\end{corollary}
These curve segments together with their endpoints are called the
{\bf spatial cylindrical curve segments (SCCS)} of $\S$ {\bf lifted
from} $e$.

To determine the edges of the topology polyhedron, an SCCS with end
points $P_1$ and $P_2$ is represented by the line segment
$e=(P_1,P_2)$. $\SE$ is the set of these line segments. For an edge
$E\in\SE$, we use $S(E)$ to denote the corresponding SCCS of $\S$.

\begin{figure}[ht]
\centering
\begin{minipage}{0.28\textwidth}
\centering
\includegraphics[scale=0.14]{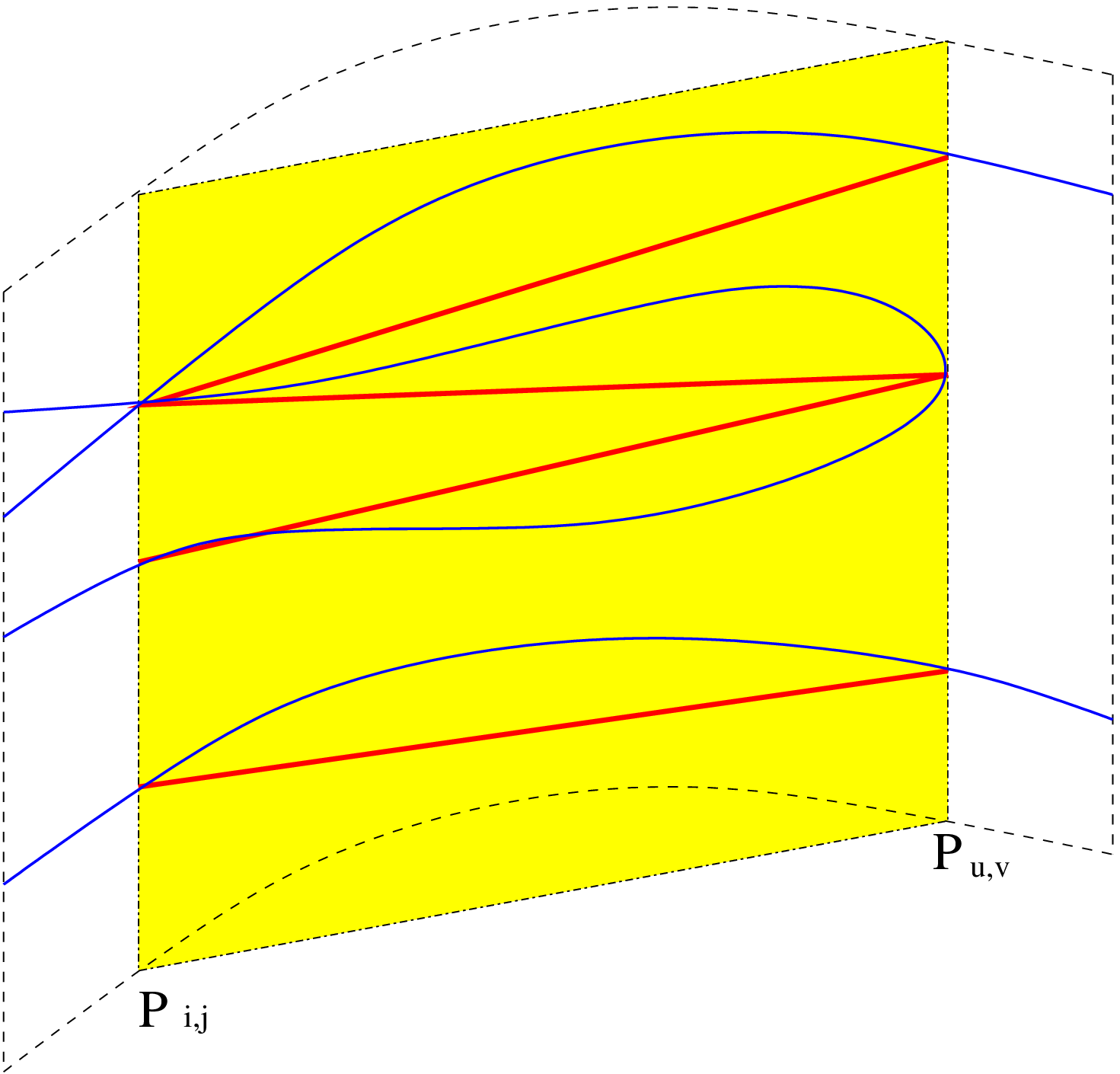}
\caption{Mesh SCCSes} \label{fig-top-spcur}
\end{minipage}\quad
\begin{minipage}{0.30\textwidth}
\centering
\includegraphics[scale=0.12]{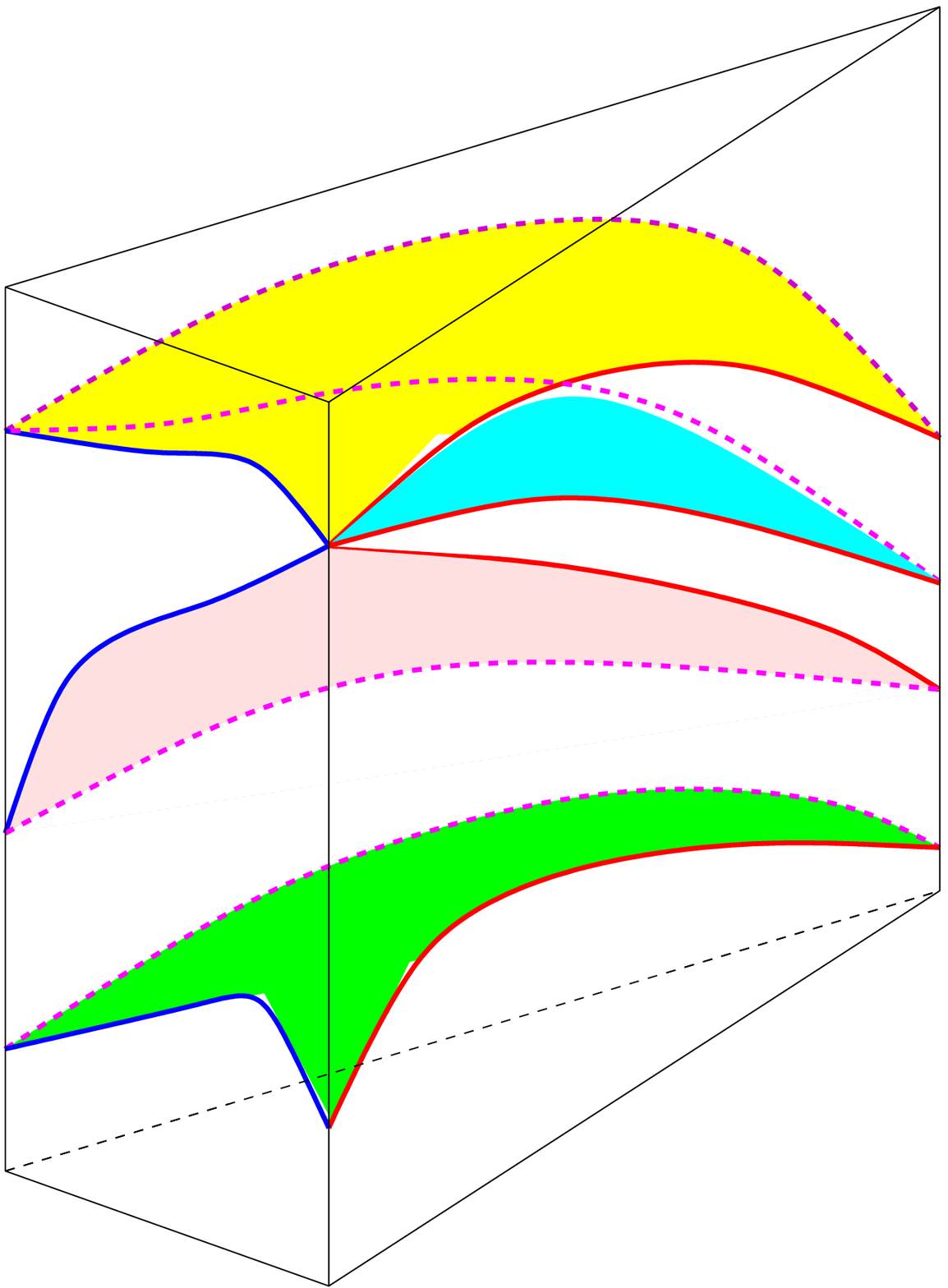}\quad
\includegraphics[scale=0.12]{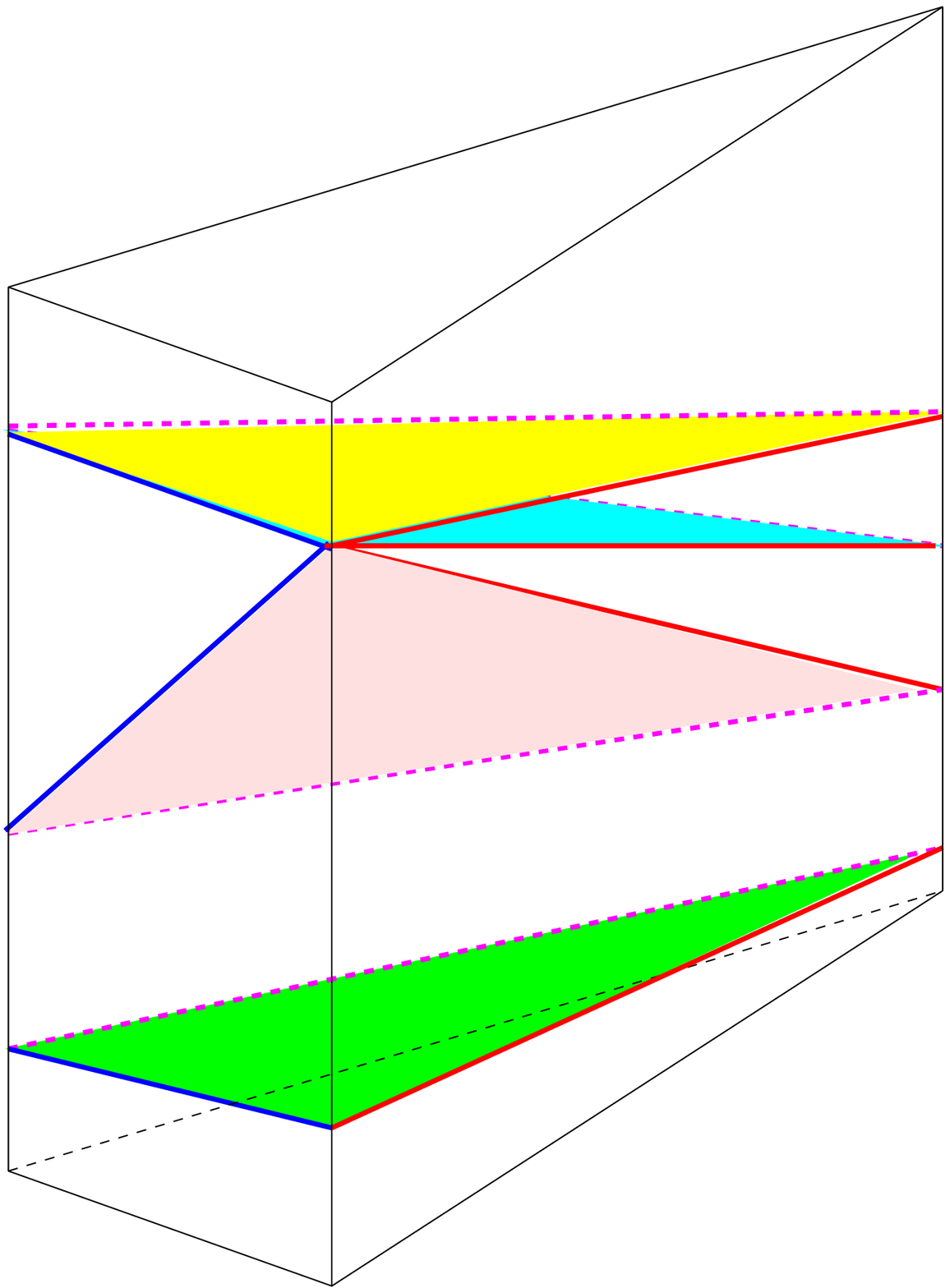}
\caption{Mesh TSPs} \label{fig-top-surfp}
\end{minipage}
\end{figure}

Let $P_{i, j, k}\in \SP$ and $e=(P_{i,j},P_{u,v})\in \EE$. We use
${\#(P_{i, j, k}, e)}$ to represent the {\bf number of SCCSes} which
have $P_{i, j, k}$ as an end point and are lifted from $C(e)$. We
use {$\#(e)$} to denote the {\bf number of SCCSes lifted from $e$}.
Define $\#(P_{u, v, w}, e)$ similarly. As a direct consequence of
Lemma \ref{lm-3c}, the following equation
\begin{eqnarray}\label{eq-numsccs}
&& \#(e)=\sum_k \#(P_{i, j, k}, e)=\sum_{w} \#(P_{u, v, w}, e)
\end{eqnarray}
holds for each $e=(P_{i, j}, P_{u, v})\in\EE$. (See Fig.
\ref{fig-top-spcur})

In  Step {\bf S4},  we find the surface patches lifted from a
triangular cell $c\in\EC$ by identifying their boundaries which are
SCCSes of $\S$. As a consequence of Theorem \ref{th-mcc}, we have

\begin{lemma}\label{lm-3s}
Let $c\in\EC$. Then $f(x,y,z)=0$ is delineable over $S=I(c)$.
\end{lemma}
{\em Proof.} For any $P=(\alpha,\beta)\in S$, $f$ is
degree-invariant and does not vanish. The discriminant $D(x,y)$ of
$f$ does not vanish on $P$. So $D$ is order-invariant over $S$. By
Theorem \ref{th-mcc}, the lemma holds. \qed

\begin{lemma}\label{lm-3s1}
$\S\cap (I(c)\times[\ZB_1, \ZB_2])$ consists of disjoint surface
patches whose edges are  SCCSes  and whose vertices are points in
$\SP$. These surface patches with their edges and vertices are
called {\bf triangular surface patches} (TSP) lifted from $c$.
\end{lemma}
{\em Proof.} By Lemma \ref{lm-3s}, the intersection of $\S$ and
$I(c)\times[\ZB_1, \ZB_2]$ consists of disjoint surface patches.
The edges of a surface patch $s$ are the intersection of $\S$ and
$I(e_i)\times[\ZB_1, \ZB_2],i=1,2,3$, where $e_i$ are the three
sides of $c$. As a consequence, the edges of these surface patches
are SCCSes.
If $c=(P_1,P_2,P_3)$, the vertices of an intersection surface patch
are the intersection points of $\S$ and $P_i\times[\ZB_1,
\ZB_2],i=1,2,3$. As a consequence, the vertices $Q_1,Q_2,Q_3$ of a
triangular surface patch are points in $\SP$ lifted from
$P_1,P_2,P_3$ respectively.\qed

It is clear that the TSPs are the intersection of $C(e)\times[\ZB_1,
\ZB_2]$ and $\S$.

For a cell $c=(P_{i, j}, P_{u, v}, P_{s, t})\in\EC$ and an edge
$E=(P_{i, j,k}, P_{u, v,w})\in\SE$ lifted from the side $e=(P_{i,
j}, P_{u, v})$ of $c$,  we use $\#(c)$ to denote the {\bf branch
number} of TSPs lifted from $R(c)$ and ${\#(E, c)}$ to denote the
{\bf number of TSPs} which pass through $S(E)$ and lifted from
$R(c)$.
Notations $\#((P_{u, v, w}, P_{s, t, l}), c)$ and $\#((P_{i, j, k},
P_{s, t, l}), c)$ can be similarly defined.
As a consequence of Lemma \ref{lm-3s1}, for $c=(P_{i, j}, P_{u, v},
P_{s, t})\in\EC$, we have
\begin{equation}\label{eq-numtcs}
\#(c)=\sum_{E_1} \#(E_1, c)=\sum_{E_2} \#(E_2, c)=\sum_{E_3} \#(E_3,
c),
\end{equation}
 where $E_1=(P_{i, j, k_1}, P_{u, v, k_2})$,
 $E_2=(P_{u, v, k_2}, P_{s, t, k_3})$,  $E_3=(P_{i, j, k_1} , P_{s, t, k_3})$ for all
 possible $k_1,k_2,k_3$. (See Fig. \ref{fig-top-surfp})

\subsection{The algorithm}

Following the analysis in the preceding section, we now give the
algorithm to construct a topology polyhedron for a given surface.

\begin{alg}\label{alg-topsur}
{\bf TopSur}$(f(x, y, z), \B_3)$.  $\S: f(x, y, z)=0$  is the
surface satisfying condition \bref{eq-co1} and $f$ is square free.
$\B_3$ is defined in \bref{eq-b3}. Output  {\bf an isotopic topology
polyhedron} $\P$ for $\S_{\B_3}$.
\end{alg}

\begin{enumerate}
\item
 {\bf Compute the projection curve} $\C: G(x,y)=0$ as in
\bref{eq-projcur}.

\item
{\bf Compute the extended topology graph}: $\EG= \{\EP$, $\EE$,
$\EC\}$ of $\C_{\B_2}$ with Algorithm \ref{alg-etopcur}, where
$\B_2$ is defined in \bref{eq-b2}.

\item
{\bf Compute $\SP$.}
For any $P_{i,j}\in\EP$, use Algorithm \ref{alg-segboxp3} with input
$(f,\B_3,P_{i,j},1)$ to compute $P_{i,j,k}$.

\item
{\bf  Compute $\SE$.} Let $\SE=\emptyset$.
\begin{enumerate}

\item
For each $P_{s,t}\in\EP$ and $e\in\EE$ with $P_{s,t}$ as an
endpoint, use Algorithm \ref{alg-edge} to compute $\#(P_{s, t, k},
e)$.

\item
For any $e=(P_{i,j},P_{u,v})\in\EE$, let $L_1=(P_{i, j, 0},$
$\ldots$, $P_{i, j, s_{i, j}})$ such that point $P_{i, j, k}$
repeats $\#(P_{i, j, k}, e)$ times.
Similarly, define $L_2=(P_{u, v, 0}, \ldots, P_{u, v, s_{u, v}})$.

\item By \bref{eq-numsccs},  $|L_1|=|L_2|=m$. Let $L_1=(P_1, \ldots, P_m)$
and $L_2=(Q_1, \ldots, Q_m)$. Add $(P_i, Q_i)$ to $\SE$.
See Fig. \ref{fig-top-spcur} for an illustration.
\end{enumerate}

\item
{\bf  Compute $\SF$.} Let $\SF=\emptyset$.
\begin{enumerate}
\item For each cell $c\in\EC$ and  $E\in\SE$
lifted from a side of $c$,  compute $\#(E, c)$ with Algorithm
\ref{alg-face}.

\item
Let $e_1,e_2,e_3\in\EE$ be the three sides of $c$.
Let $S_i$ be the sequence of edges in $\SE$ lifted from $e_i$
ordered  bottom up and an $E$ is repeated $\#(E, c)$ times in the
sequence.

\item By \bref{eq-numtcs}, $|S_1|=|S_2|=|S_3|=t$.
Let $S_i = \{ E_{i,k}$, $k=1, \ldots, t\}$. Then the three line
segments $E_{1,k},E_{2,k},E_{3,k}$ should form a triangle
$f=(P_{1,k},P_{2,k},P_{3,k})$. Add $f$ to $\SF$. See Fig.
\ref{fig-top-surfp} for an illustration.

\end{enumerate}

\item Output $\P = \{\SP, \SE,
\SF\}$. The isotopic map can be computed as usual \cite{pv2}.
\end{enumerate}

\begin{theorem}\label{th-s1}
Algorithm \ref{alg-topsur} computes an isotopic meshing for
$\S_{\mathbf{\B_3}}$.
\end{theorem}
{\em Proof. } First, we prove the algorithm compute the correct
topology of given surface. Note that with the auxiliary points added
in Step 6(a) of Algorithm \ref{alg-topcur} and Step 3 of Algorithm
\ref{alg-etopcur}, the edges in Step 4(c) and the faces in Step 5(d)
are mutually different. Thus, we have a well-defined polyhedron.

The extended topology graph $EG$ divides the rectangle
$\mathbf{B}_2$ into triangular cells. We need only to show that for
each edge $e\in\EE$ and each cell $c\in\EC$,  $\P$ and $\S$ have the
same topology on $C(e)\times [\ZB_1, \ZB_2]$ and $C(c)\times [\ZB_1,
\ZB_2]$ respectively.

For $e\in\EE$, from Step 4  the SCCSes of $\S$ on the cylindrical
surface $S_1=C(e)\times [\ZB_1, \ZB_2]$ do not intersect except at
the end points. By Corollaries \ref{cor-3c1} and \bref{eq-numsccs},
the edges of $\SE$ are the line segments with the same end points as
those SCCSes. Then, the plane graph $\P$ on $e\times[\ZB_1, \ZB_2]$
and $\S$ on $S_1$ have the same topology. See Figure
\ref{fig-top-spcur} for an illustration. The spatial curve segments
are presented by line segments.
With similar arguments,  we could show that the part of $\P$ on
$c\times [\ZB_1, \ZB_2]$ and $\S$ on $C(c)\times [\ZB_1, \ZB_2]$
have the same topology. See Figure \ref{fig-top-surfp} for an
illustration.
This proves the topology correctness of the algorithm.

Then we prove the topology polyhedron is a isotopic meshing of the
given surface.

The extended topology graph $ \EG=\{\EP,\EE,\EC\}$ for the curve
$\C_{\B_2}$ decompose $\B_2$ into triangular cells.
According to Theorem \ref{th-c1}, $\EG$ and $\C$ are isotopic and we
can construct a homeomorphism $F$ from $\R^2$ to itself that deforms
$\EE$ to $\C$ continuously:
$$F: \R^2\times[0,1]\rightarrow\R^2.$$

Let  $\P=(\SP,\SE,\SF)$ be a topology polyhedron for a surface
$\S_{\mathbf{\B_3}}$ which decomposes $\B_3$ into cylindrical
regions in a similar way as described in the proof of Theorem
\ref{th-c1}. Extend $F$ to $\R^3\times[0,1]$:
$$T_1=(F(x,y),z): \R^3\times[0,1]\rightarrow\R^3.$$
The inverse transformation $T_1^{-1}$ of $T_1$ deforms all SCCSs of
$\S$ into planes  $\{C(e)\times\R, e\in \EE\}$ which are
perpendicular to the $xy$-pane. Denote $\S_1$ to be the surface
$T_1^{-1}(\S)$. We need only to prove that $\S_1$ and $\P$ are
isotopic.

We can construct a homeomorphism  $T_2$ from $\R^3$ to itself
similar to that give in the proof of Theorem \ref{th-c1} to deform
the $z$ direction such that $T_2(\SF,0)=\SF$ and $T_2(\SF,1)=\S_1$.

The transformation $T=T_1\circ T_2$ is a homeomorphism from $\R^3$
to itself which deforms $\SF$ to $\S$ continuously. \qed

We implemented Algorithm \ref{alg-topsur} in Maple. Two groups of
experiments are done for the following five surfaces with
singularities.

{\footnotesize
\parindent=0pt\parskip=2pt
$\S_1:f_1=x^4+y^4+z^4-x^2-y^2-z^2-x^2y^2-x^2z^2-y^2z^2+1=0$, $\B_3 =
[[-1.5,1.5],[-1.25,1.25],[-2,2]]$.

$\S_2:f_2 =
-1+(27/2)z^2y^2x^2-(27/2)x^2y^2-6x^2z^2-(27/2)y^2z^2+3x^2+3z^2+
(27/4)y^2-3x^4-(243/16)y^4-3z^4+x^6+(729/64)y^6+z^6+
(27/4)x^4y^2+3x^4z^2+(243/16)x^2y^4+3x^2z^4+(243/16)z^2y^4+
(27/4)z^4y^2-x^2z^3-(9/80)y^2z^3=0$, $\B_3 =[[-2,2],[-2,2],[-4,4]]$.

$\S_3:f_3=-2y^4+2y^2z^2+y^2+z^4-2z^2+x^6+3x^4y^2-3x^4+3x^2y^4-6x^2y^2+3x^2+y^6=0$,
$\B_3 =[[-2,2],[-2,2],[-2,2]]$.

$\S_4:f_4=x^2y^2+y^2z^2+z^2x^2-7xyz/2=0$, $\B_3
=[[-2,2],[-2,2],[-2,2]]$.

$\S_5:f_5=16-2x^2z^2-8z^2+4x^3-x^5+(1/4)x^6+x^4+y^4+y^2x^3+z^4+z^2x^3-2x^2y^2+2y^2z^2-8x^2-8y^2=0$,
$\B_3 =[[-2,2],[-3,3],[-6,6]]$. }

\begin{figure}[ht]
\begin{minipage}{0.555\textwidth}
\centering
\includegraphics[scale=0.33]{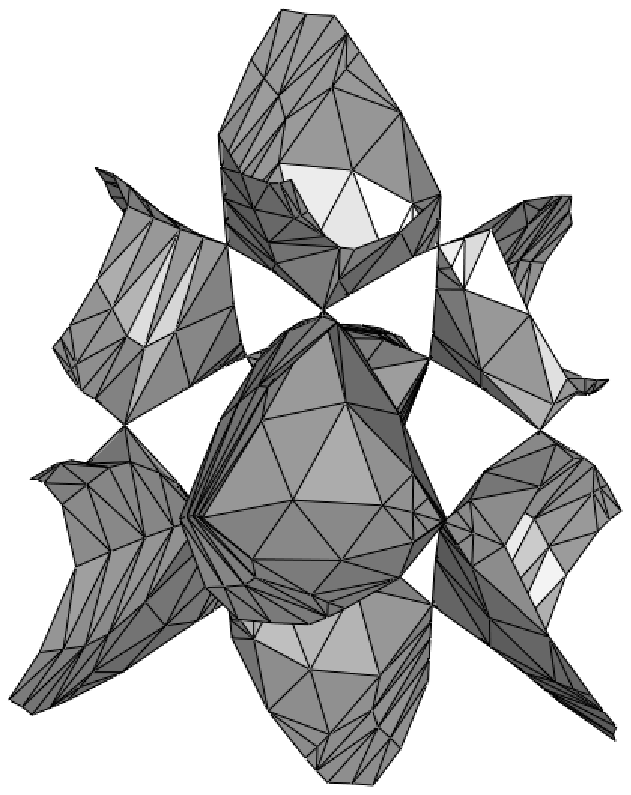}\quad
\includegraphics[scale=0.70]{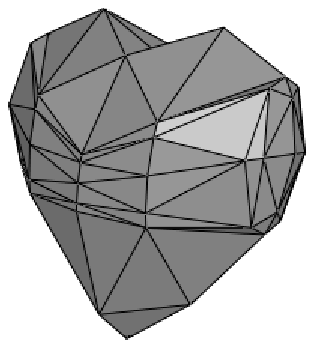}
\caption{Topology polyhedrons for surfaces $\S_1$ and $\S_2$.
}\label{fig-tops}
\end{minipage}
\begin{minipage}{0.335\textwidth}
\centering
\includegraphics[scale=0.78]{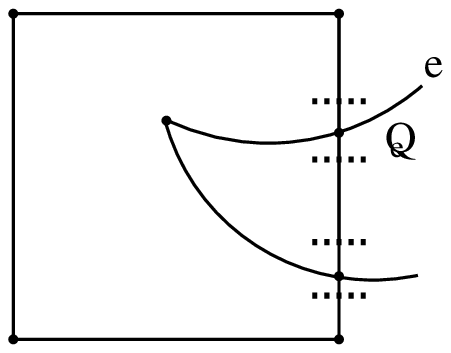}
\caption{Isolation intervals}\label{fig-qe}
\end{minipage}
\end{figure}

The first experiment is to compute an isotopic polyhedron for the
surfaces without considering precision. The timings are given in the
second row of Table \ref{tab}. Two of the polyhedrons are shown in
Fig. \ref{fig-tops}.
In the second experiment, we continue to subdivide the intervals
between $[\XB_1,\XB_2]$ to compute a more accurate meshing. The
results are given in Fig. 1. The timings are given in the third row
of Table \ref{tab}. $\#$Mesh in the fourth row gives the number of
meshes in these meshings.
Considering that implementations in Maple are generally slow due to
overhead costs, our algorithm is quite effective.

\begin{table}[h]
\centering
  \begin{tabular}{|c|c|c|c|c|c|}
    \hline
     TYPE & $\mathcal{S}_1$ & $\mathcal{S}_2$ & $\mathcal{S}_3$ & $\mathcal{S}_4$& $\mathcal{S}_5$  \\
     \hline
     Topology& 0.544  & 0.816 & 0.760  &0.684&1.280\\
     \hline
     Meshing & 11.7 & 11.8 & 22.0 &51.1& 92.0\\
     \hline
     $\#$Mesh& 1472 & 1612 & 3032 &3658 & 5456\\
     \hline
  \end{tabular}\\
\caption{Timings on a PC with Linux OS, 3.00G Core 2Duo CPU, and 2G
RAM. } \label{tab}
\end{table}

\subsection{Segregating box for a point on $\S$}\label{sec-segboxp}

Assume that $\SP$ is of form \bref{eq-ps3}. Then $\B_{i,j}$ in
\bref{eq-ps2} is an isolation box for $P_{i,j}$ and $\B_{i,j,k}$ is
an isolation box for $P_{i,j,k}$.
It is clear that
 \begin{equation}\label{eq-c1}
 f(\alpha_i,\beta_{i,j},e_{i,j,k})f(\alpha_i,\beta_{i,j},d_{i,j,k})\ne0
 \end{equation}
The isolating box $\B_{i,j,k}$ of $P_{i,j,k}$ is called a {\bf
segregating box} if $f(x,y,z)$ does not intersect with the top and
bottom faces of $\B_{i,j,k}$.
Due to \bref{eq-c1}, when sufficiently subdividing $\B_{i,j}$,
$\B_{i,j,k}$ will become a segregating box. This leads to the
following algorithm.

\begin{alg}\label{alg-segboxp3}{\bf
SegBoxP3}$(f(x,y,z),\B_3,P,\epsilon)$ where $\S$: $f(x, y, z)$ $=0$
is the surface, $\B_3$ defined in \bref{eq-b3}, $P$ a plane point
defined by  $\Sigma_2=\{h(x),g(x,y)\}$ and an isolation box $\B$,
and $\epsilon>0$.
Output the set of points $\{P_i\}$ on $\S$ lifted from $P$,
segregating boxes for $P_i$, and a new segregating box $\B$ of $P$.
\end{alg}
\begin{enumerate}

\item
Let $\{\B_1,\ldots,\B_s\}=\ROOTI(\Sigma_3,\B\times [\ZB_1,
\ZB_2],\epsilon)$, where $\Sigma_3 = \{h(x),g(x,y),f(x,y,z)\}$.

\item Let $\B_{i}=\B\times[e_{i},f_{i}]$
be the isolation box for $P_{i}$ on $\S$.

\item Let $\eta=\epsilon$. While $0\in\intbox f(\B\times[e_{i},e_i])$ or $0\in\intbox
f(\B\times[f_{i},f_{i}])$ for some $k\in \{1,\ldots,s\}$, repeat

\hskip 1truecm $\eta = \eta/2$ and  $\B:=\ROOTI(\Sigma_2, \B,
\eta)$.


\item
Output the points $P_i$ defined by $\Sigma_3$ and $\B_i$, and the
new $\B$.

\end{enumerate}

In Step 3, if $f(\alpha_i,\beta_{i,j},\YB_1)=0$ or
$f(\alpha_i,\beta_{i,j},\YB_2)=0$, then we need to use the minimal
circle method introduced in \cite{chengtop} to find a segregating
box in order for Lemma \ref{lm-sn} to be true at this boundary
point.

\subsection{Compute number of SCCSes adjacent to a point }
\label{sec-edge}

Let $P_{i,j,k}$ be a point lifted from point $P_{i,j}$ and
$e=(P_{i,j},P_{u,v})\in\EE$. We will show how to compute $\#(P_{i,
j, k}, e)$.

For any point $P$ on the projection curve $\C: G(x,y)=0$ and a
segregating box $\B=[a,b]\times[c,d]$ of $P$,
$\C$  intersects only with the vertical boundaries of $\B$.

\begin{figure}[ht]
\centering
\begin{minipage}{0.34\textwidth}
\centering
\includegraphics[scale=0.17]{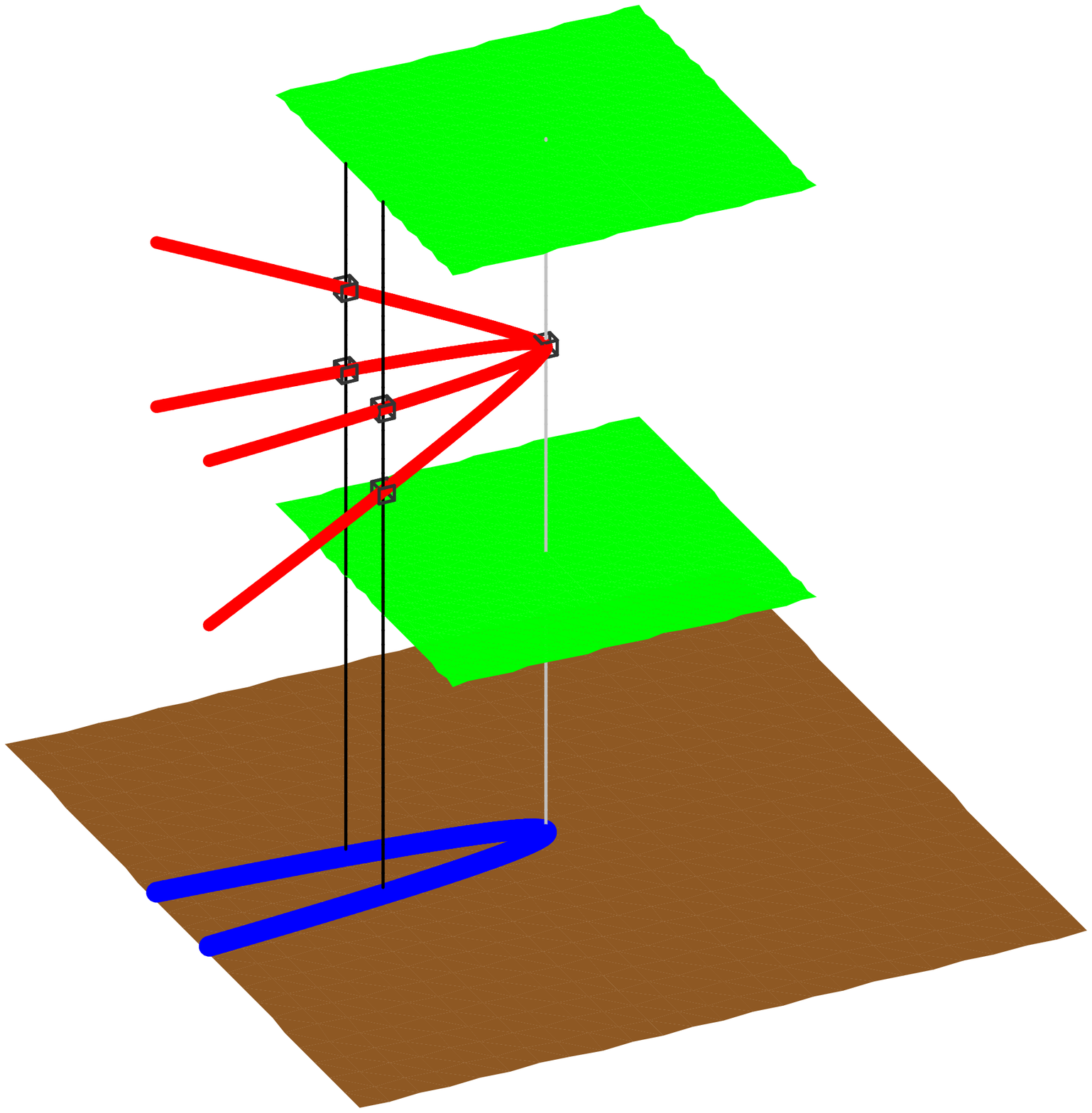}
 \caption{Compute $\#(P_{i,j,k},e)$}
\label{fig-num-spcurve}
\end{minipage}\quad
\begin{minipage}{0.29\textwidth}
\centering
\includegraphics[scale=0.17]{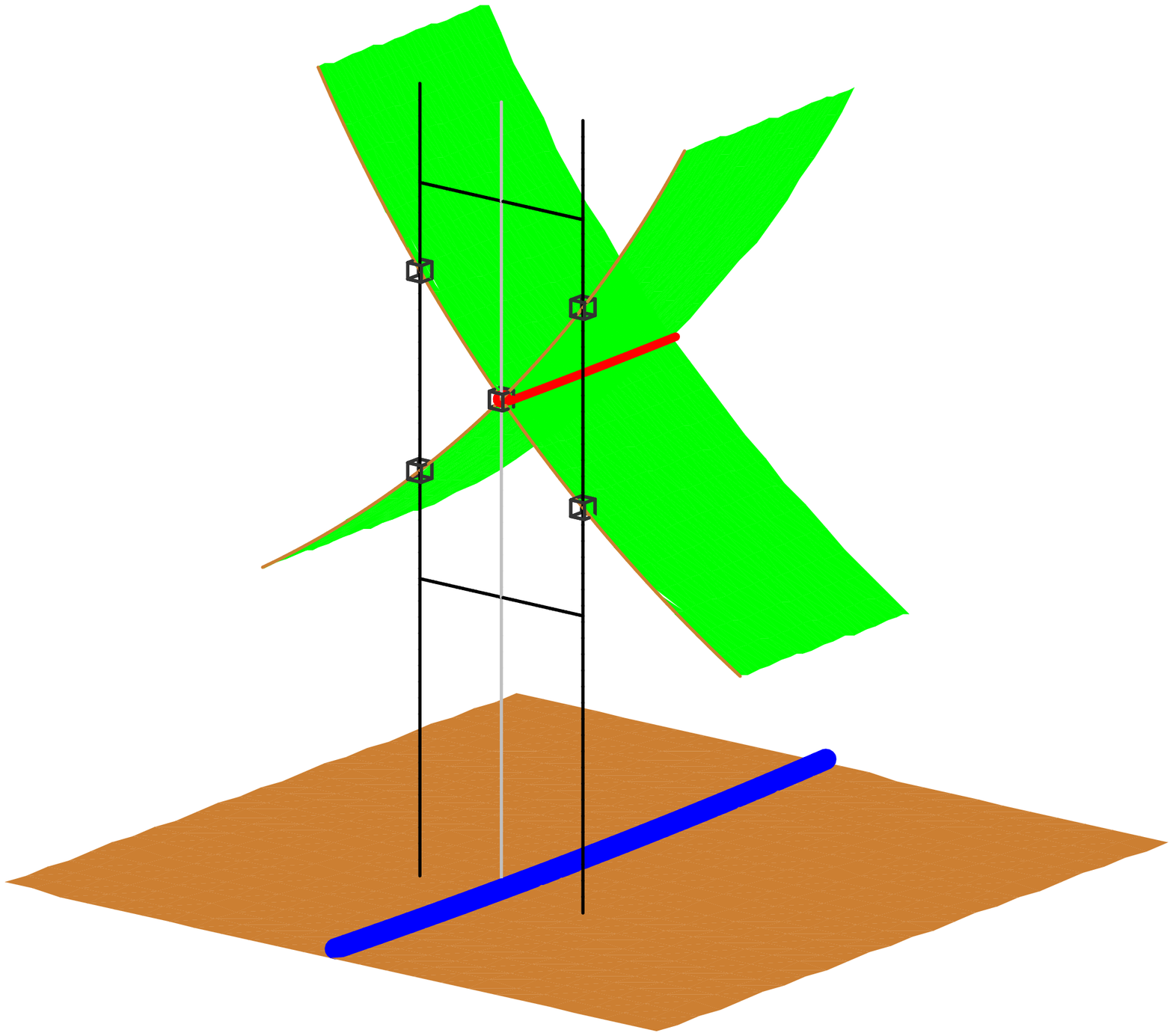}
 \caption{Compute $\#(S,c)$} \label{fig-striangle}
\end{minipage}
\end{figure}

For an edge $e\in\PE$, consider the right boundaries of $\B$. We
denote the intersection point of $C(e)$ with line $x=b$ as $Q_e$ and
\begin{equation}\label{eq-c2}[b,b]\times[u_e,v_e]\end{equation} is an isolation interval for $Q_e$ on line $x=b$, which  is called
the {\bf isolation interval} of $C(e)$. See Fig. \ref{fig-qe}.
\begin{lemma}\label{lm-sn}
Use the above notations. If $\B_{i,j,k}$ is a segregating box for
$P_{i,j,k}$ and $\S$ is delineable over $I(e)$,
then $\#(P_{i,j,k},e)$ equals to the number of solutions of the
triangular system $\Sigma_R=\{G(b_i,y),f(b_i,y,z)\}$ in the interval
box $[u_e,v_e]\times[e_{i,j,k},f_{i,j,k}]$. Geometrically, this is
the number of intersection points of the line segment
$\{x=b_i,y=\gamma_i, e_{i,j,k}\le z \le f_{i,j,k} \}$ and the
surface $\S$ where $(b_i,\gamma_i)$ is a point on $G(b_i,y)=0$. See
Fig. \ref{fig-num-spcurve} for an illustration.
\end{lemma}
{\em Proof.} From Algorithm \ref{alg-branch}, each SCCS passing
through $P_{i,j,k}$ and projecting to $C(e)$ must pass through the
rectangle $[b_i,b_i]\times[u_e,v_e]\times[\ZB_1,\ZB_2]$. Since
$\B_{i,j,k}$ is a segregating box, these SCCSes must intersect with
the the rectangle
$\RR=[b_i,b_i]\times[u_e,v_e]\times[e_{i,j,k},f_{i,j,k}]$. Further,
each SCCS can intersect with the rectangle only once since these
SCCSes are delineable by  Lemma \ref{lm-3c}. Note that the number of
solutions of the triangular system $\Sigma_R$ is the number of
intersections of the SCCSes and the rectangle $\RR$.\qed

{\bf Remark.} Similarly, we can compute the number of the SCCSes on
the left hand side of the point $P_{i,j,k}$ by computing the number
of solutions for $\{G(a_i,y)=0,f(a_i,y,z)=0\}$. When $G(x,y)=0$
contains vertical lines, we can compute the number of SCCSes passing
through $P_{i,j,k}$ and projecting to these lines by solving
$\{G(x,w),f(x,w,z)\}$ for $w=c_{i,j}$ and $w=d_{i,j}$ respectively.

We now give the following algorithm to compute the number of curve
branches.

The following algorithm is based on  Lemma \ref{lm-sn}.

\begin{alg}\label{alg-edge}{\bf
NumSCCS}$(f(x,y,z),P_{i,j,k},e)$  $\S:f(x, y,z)$ $=0$ is a surface
delineable over $I(e)$, $P_{i,j,k}\in\SP$ is of form \bref{eq-ps3},
and $e\in\EE$ is an edge with $P_{i,j}$ as an end point, where
$P_{i,j}$ is the projection point of $P_{i,j,k}$. The output is
$\#(P_{i,j,k}, e)$.
\end{alg}
\begin{enumerate}

\item If $e$ is an $x$-vertical line segment above $P_{i,j}$ in the $y$-direction,
then form the triangular system $\Sigma_{22}=\{g_i(x,d_{i,j})$,
$f(x, d_{i,j}, z)\}$ and let $\QQ=\ROOTI(\Sigma_{22}, [a_i,
b_i]\times[e_{i,j,k}, f_{i,j,k}], 1)$. Output  $\#(P_{i,j,k}, e)=
|\QQ|$.

\item If $e$ is not an $x$-vertical line segment, we need to
compute the isolation intervals defined in \bref{eq-c2}.
We only consider the right branches.
Let $\RR=\ROOTI(g_i(b_i, y),[c_{i, j}, d_{i, j}],1)$ where
$r=R\#(P_{i, j})=|\RR|$.
By Lemma \ref{lem-branch}, one of intervals in $\RR$ is the
isolation interval $[u_e,v_e]$ for $e$.

\item Let $\Sigma_{21}=\{g_i(b_i, y), f(b_i, y, z)\}$ be a triangular
system in $y$ and $z$ and $\QQ=\ROOTI(\Sigma_{21}$,
$[u_e,v_e]\times[e_{i,j,k}, f_{i,j,k}], 1)$.
Output $\#(P_{i,j,k}, e)= |\QQ|$. See Fig. \ref{fig-num-spcurve} for
an illustration.

\end{enumerate}

If there exist no SCCSes originating from a point, it is an {\it
isolated singularity}.

\subsection{Compute number of TSPs adjacent to an SCCS}
\label{sec-face}

We  compute the number of TSPs originating from an $E\in\SE$. That
is, for an $E=(P_{i,j,k},P_{u,v,w})\in\SE$ and a $c\in\EC$ with
$e=(P_{i,j},P_{u,v})$ as an edge, we will compute $\#(E,c)$.

Use the notations in Algorithm \ref{alg-edge}. Denote the SCCSs
passing through point $P_{i,j,k}$ and projecting to $C(e)$ as
$S(s_i),i=1,\ldots,m$. Assume that $\QQ$ (Step 3 of Algorithm
\ref{alg-edge}) is the set of isolation boxes of $m$ points
$Q_1,\ldots,Q_s$ with $Q_i$ on $S(s_i)$.
Then in the plane $x=b_i$ (or $x=a_i$), the surface becomes a plane
curve $f(b_i, y, z)=0$ and each surface patch passing through
$S(s_i)$ becomes a curve segment of the curve $f(b_i, y, z)=0$
passing through $Q_i$. We summarize this as the following lemma.

\begin{lemma}\label{lm-numtsp}
Use the above notations. If $\S$ is delineable over $I(e)$ and
$I(c)$ respectively, then the number of TSPs passing through
$S(s_i)$ and projecting to $R(c)$ is the number of curve branches
passing through $Q_i$ and projecting to the region $R(c)$.
\end{lemma}

According to the above discussion, we have the following algorithm.

\begin{alg}\label{alg-face}{\bf
NumTSP}$(f(x,y,z),P_{i,j,k},e,c)$ $\S:f(x, y, z)=0$ is the surface
delineable over $I(e)$ and $I(c)$ respectively, $P_{i,j,k}\in\SP$,
$e=(P_{i,j},P_{u,v})\in\EE$, and $c\in\EC$ with $e$ as an edge. The
output is $\#(E_i,c)$ where $S(E_i)$ are all the SCCSes passing
through $P_{i,j,k}$ and projecting to $C(e)$.
\end{alg}
\begin{enumerate}

\item Execute Algorithm {\bf
NumSCCS}$(f(x,y,z),P_{i,j,k},e)$.

\item If  $e$ is not an $x$-vertical edge, execute the following steps
\begin{enumerate}
\item
Let  $\QQ=\{Q_1,\ldots,Q_m\}$ be the points obtained in Step 3 of
Algorithm {\bf NumSCCS}.

\item Let $\Sigma_{21}=\{g(b_i, y), f(b_i, y, z)\}$ be the defining
triangular system for $\QQ$. Execute Algorithm \ref{alg-branch} with
input $\QQ$ to compute $L\#(Q_i)$ and $R\#(Q_i)$.

\item
Let $c_1$ be the cell under $e$ in the $y$ direction and $c_2$ the
one above $e$. By Lemma \ref{lm-numtsp}, $\#(S_l, c_1) = L\#(Q_i)$
and $\#(S_l, c_2) = R\#(Q_i)$.
See Fig. \ref{fig-striangle} for an illustration.

\end{enumerate}

\item If $e$ is an $x$-vertical edge, execute the following steps
\begin{enumerate}
\item
Let $\RR=\{R_1,\ldots,R_s\}$ be the points obtained in Step 1 of
Algorithm {\bf NumSCCS}.

\item Let $\Sigma_{22}=\{g(x,d_{i,j}), f(x, d_{i,j}, z)\}$ be the
defining triangular system for $\RR$.
Execute Algorithm \ref{alg-branch}  with input $\RR$.

\item Let $c_1$ be the  cell on the left hand side of $e$ and and
$c_2$ the  cell on the right hand side of $e$. By Lemma
\ref{lm-numtsp}, $\#(T_l, c_1) = L\#(R_l)$ and $\#(T_l, c_2) =
R\#(R_l)$.
\end{enumerate}

\end{enumerate}

If there exist no TSPs connect to a SCCS, then the SCCS is an {\it
isolated spatial curve segment}.

\subsection{The General Case} \label{sec-gen}

Until now, we assume that the surface $\S$ does not contain straight
lines parallel to the $z$-axis. In this subsection, we will show how
to treat surfaces that contain such lines.

The aim is to get the points on the vertical lines where the
topology of the surface changed, and the intersections between some
SCCSes and the vertical lines, then the SCCSes originating from
these points, and the surface patches originating from the line
segments defined by these points.

The following will show how to compute the special case when
$f(\alpha, \beta, z)\equiv0$ for some point $P=(\alpha, \beta)$. It
is clear that $g(x, y)=0$ has a finite number of such points since
$f(x,y,z)$ has no factor containing $x,y$ only. We can solve the
problem in the following way.

\begin{enumerate}
\item Take a coordinate system transformation such that the transformed line $L_1$ of the vertical
line $L_0$ can be projected as a line $L_2$ on the new $XY$-plane.
\item Determine the topological information of $L_2$: the intersections
of $L_2$ and the new projection curve,  the number of curve segments
originating from each intersection on its two sides.

\item Determine the topological information of $L_1$: lifting the
intersections of $L_2$ and the projection curve of the new surface
to determine the corresponding points on $L_1$. Find the points
where the topology of surface changed on $L_1$.

\item We can made the same coordinate system transformation for the
intersection of two surfaces  $G(x,y)=0$ and $f(x,y,z)=0$. Then we
can decide the points on the vertical lines which are the
intersections of SCCSes and the vertical line.

\item Find the points where the topology of the original surface changed on $L_0$ from $L_1$ by coordinate relationship.
Determine the topological information of $L_0$.
\end{enumerate}

\noindent{\bf Remark:} It is convenient to take a transformation
such that $L_2$ is a vertical line of the new projection curve if
$L_2$ can't overlap other line(s) of the projection curve.

In this way,  we can solve the special case in Algorithm
\ref{alg-topsur}. Since we have introduced the operations we need
before, we just use an example to show the effectivity.

In this special case,  $\SE$ contains the edge with the from
$(P_{i,j,k},P_{i,j,k+1})$. Similarly, $\SF$ contain the face with
the form $(P_{i,j,k},P_{i,j,k+1},P_{u,v,w})$.

\begin{figure}[ht]
\begin{minipage}{0.60\textwidth}
 \centering
\includegraphics[scale=0.2]{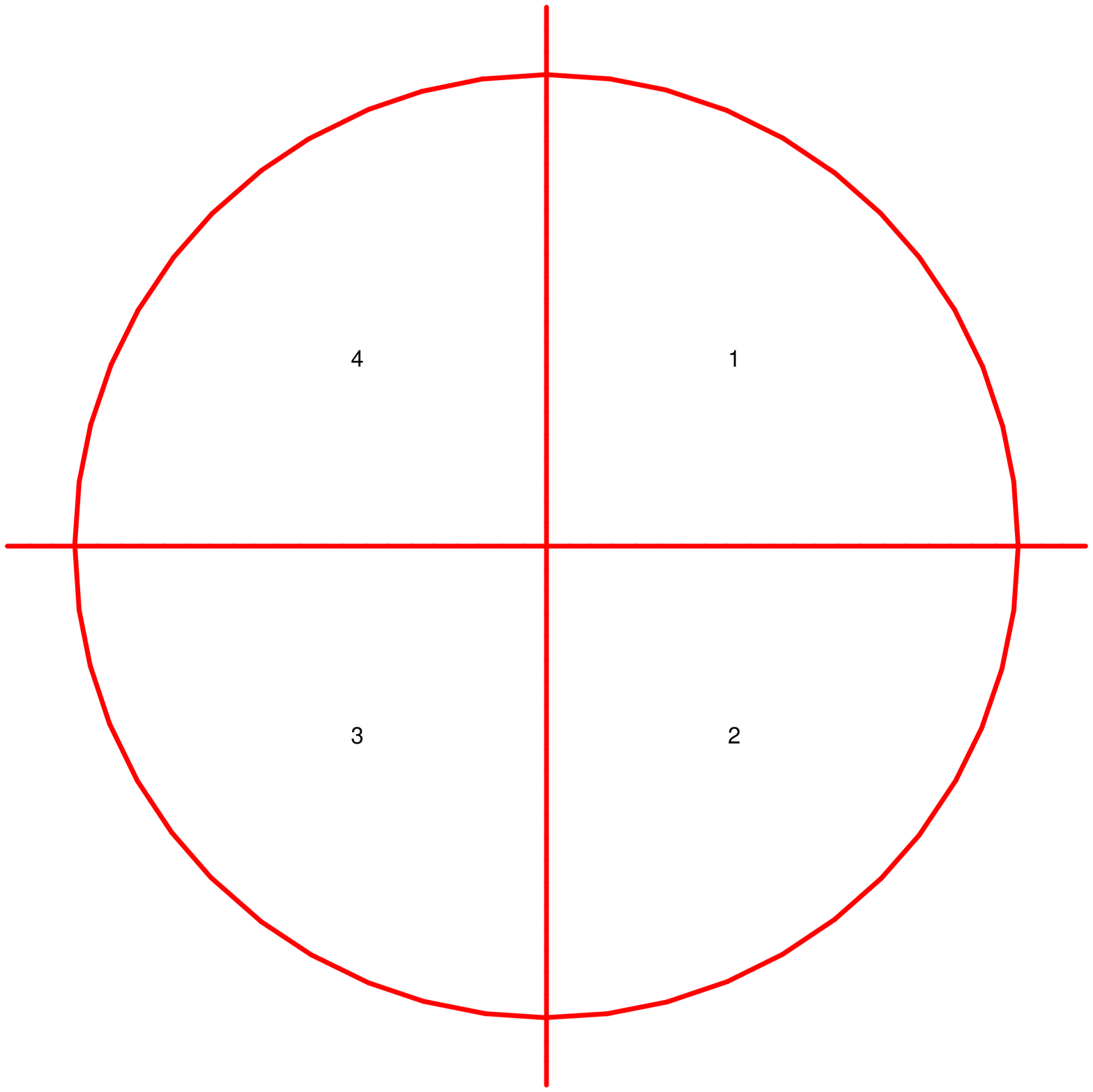}
\includegraphics[scale=0.2]{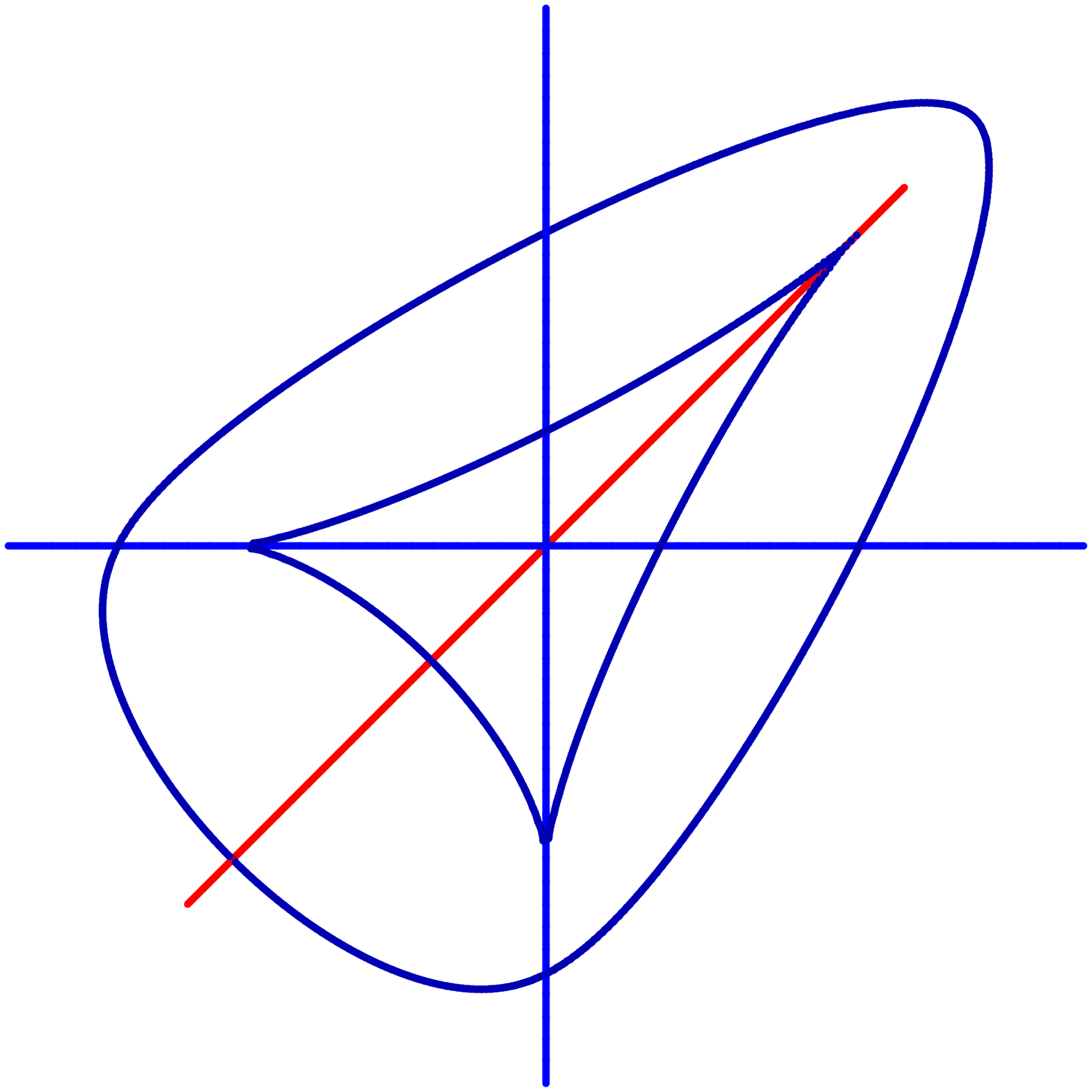}
\caption{Determining the $V_i$ in special case} \label{fig-vpoint}
\end{minipage}
\begin{minipage}{0.34\textwidth}
 \centering
\includegraphics[scale=0.2]{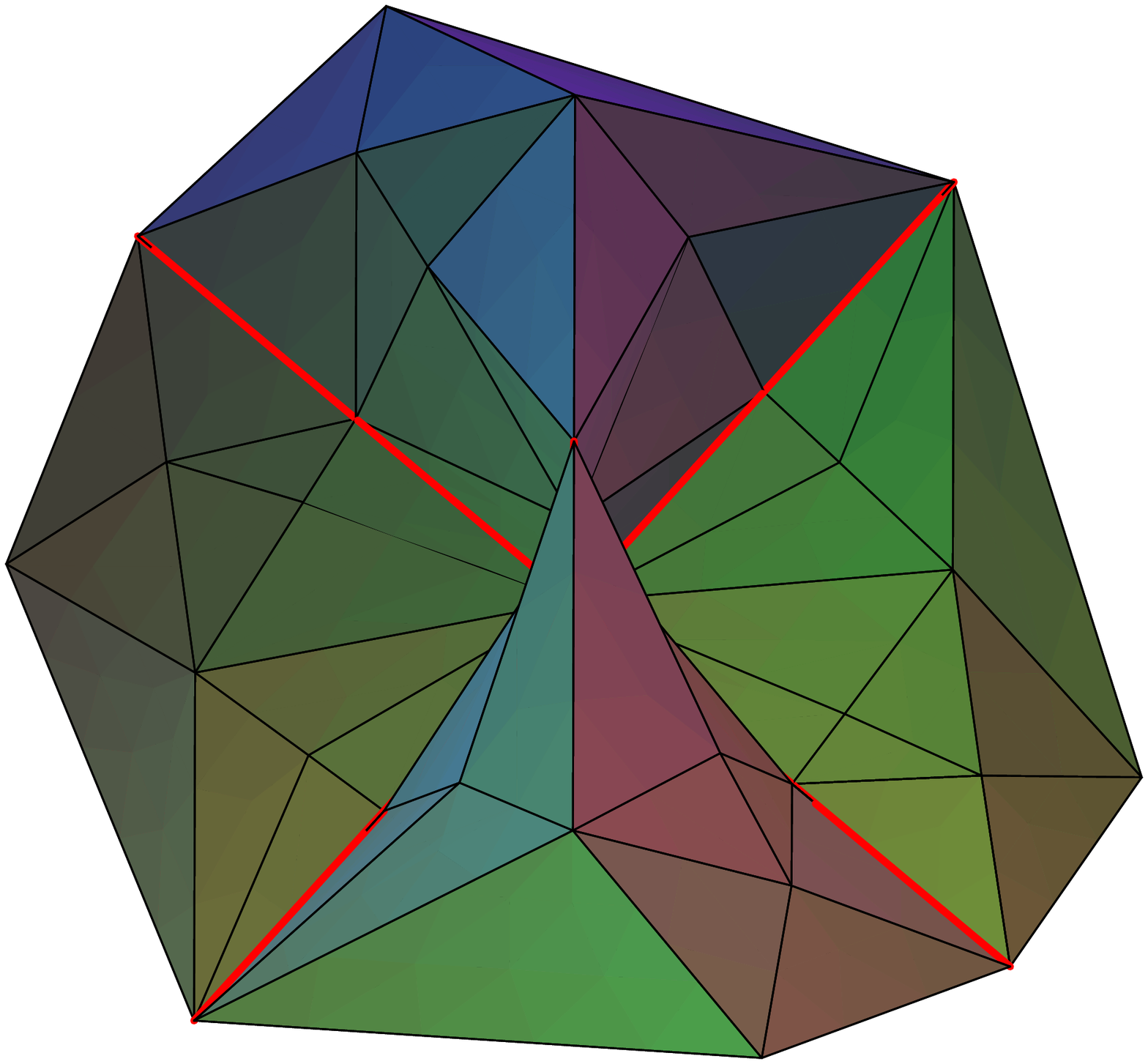}
\caption{Topology polyhedron of a surface with vertical line}
\label{fig-vertical}
\end{minipage}
\end{figure}

We will continue the same example. Let us consider the following
surface inside $\mathbb{B}=[-2, 2]\times[-2, 2]\times[-2, 2]$ as an
example.
\begin{equation}\label{eq-vertical}
\S: f(x, y, z)=x^2y^2+x^2z^2+y^2z^2-\frac{7}{2}xyz=0.
\end{equation}
It is clear that $f(0, 0, z)\equiv0$. So $(0, 0)$ is a point in
special case. So $L_0: \{x=0,y=0,-2\le z\le 2\}$. The topology
polyhedron of the surface is shown in Figure \ref{fig-vertical}.

{\bf 1.} Take the system transformation
\begin{equation}\label{equ-trans} \{x=X-Z, y=Y-Z, z=Z.\}\end{equation} We get a
new surface
\[\small
\begin{array}{lll}
\S': F(X, Y, Z)&=&{X}^{2}{Y}^{2}-2\, {X}^{2}YZ+2\,
{X}^{2}{Z}^{2}-2\, XZ{Y}^{2}+4\, X{Z}^{2} Y-4\, X{Z}^{3}+2\,
{Z}^{2}{Y}^{2}\\ &&-4\, {Z}^{3}Y+3\, {Z}^{4}-\frac{7}{2}\,
ZXY+\frac{7}{2}\, X
{Z}^{2}+\frac{7}{2}\, Y{Z}^{2}-\frac{7}{2}\, {Z}^{3}\\
&=&0.
\end{array}
\]
Now $L_0$ corresponds to the line segment $L_1: \{X=Z,Y=Z,-2\le Z\le
2\}$ on the new surface.

{\bf 2.} The projection curve of $\S'$ is shown in the right part of
Figure \ref{fig-vpoint}. The red line segment is the $L_2:
\{X-Y=0,-2\le X\le 2\}$. It corresponds to $L_1$. The isolation
boxes of the singularities of the projection curve of $\S'$ on $L_2$
are below.

$[P_1,P_2,P_3,P_4]:=[[-\frac{7}{4},
-\frac{7}{4}]\times[-\frac{7}{4}, -\frac{7}{4}], [-{\frac {41}{64}}
, -{\frac {655}{1024}}]\times[-{\frac {41}{64}} , -{\frac
{655}{1024}}], [0, 0]\times[0, 0], [\frac{7}{4},
\frac{7}{4}]\times[\frac{7}{4}, \frac{7}{4}]]. $

{\bf 3.} The corresponding points $Q_i$ on $L_1$ of these
singularities $P_1,P_2,P_3,P_4$ are
$$[Q_1,Q_2,Q_3,Q_4]:=[ [t\times t\times t],$$
$$t=[-\frac{7}{4}, -\frac{7}{4}],[-{\frac {41}{64}} , -{\frac
{655}{1024}}],[0, 0],[\frac{7}{4}, \frac{7}{4}]].$$
Assume the endpoints of $L_1$ are $Q_0,Q_5$. Computing the SCCSes
originating from $Q_i (i=0,\ldots,5)$ with Algorithm \ref{alg-edge},
we can find that:

There are no SCCSes originating from $Q_0 (Q_5)$ except for
$\overline{Q_0Q_1}$ ($\overline{Q_4Q_5}$) on $L_1$, we can find
there are on surface patches originating from $\overline{Q_0Q_1}$
($\overline{Q_4Q_5}$) by Algorithm \ref{alg-face}; Similarly, $Q_1$
originates one SCCS from $L_1$'s two sides respectively, and the
SCCSes originates two surface patches in the two cells besides
$\overline{Q_1Q_2}$, so $Q_1$ is a point we are interested; $Q_2$
originates two SCCSes from $Q_2$ besides $L_1$ respectively, and the
four SCCSes all originate one surface patch on the cells besides
$\overline{Q_1Q_2}$ and $\overline{Q_2Q_3}$, which means $Q_2$ is
not a point where the topology of the new surface changes on $L_1$;
$Q_3$ originate two line segments parallelling to $XY$-plane as
SCCSes on $L_1$'s two sides respectively, all originating 2 surface
patches on the cell bodies besides them, which means the SCCSes are
singular curve of the surface, so $Q_3$ is a point we are
interested; $Q_4$ originate one SCCS on $L_1$'s two sides
respectively, $\overline{Q_3Q_4}$ originates surface patches but
there is no surface patches originating from $\overline{Q_4Q_5}$, so
$Q_5$ is a point we are interested.

So we can conclude that $Q_1,Q_3,Q_4$ are the points where the
topology of the surface changed on $L_1$.

{\bf 4.} Take the same coordinate system transformation as
(\ref{equ-trans}) for $g=x y (16 x^2+16 y^2 -49)$, We can get a
surface:
$$ G(X,Y,Z) = (X-Z) (Y-Z) (16 (X-Z)^2+16 (Y-Z)^2 -49).$$
We just need to decide some points on the line corresponding to the
vertical line on the space curve defined by $G(X,Y,Z)=0$ and
$F(X,Y,Z)=0$.
 Use the method in \cite{mourrainsc}, we can find that there is only one point on the
vertical line which is the intersection of SCCSes and the vertical
line. It is $[0,0]\times[0,0]\times[0,0]$.

{\bf 5.} Now we can get the points where we are interested on $L_0$,
we can simply call these points as {\bf vertical points}. By the
coordinate relationship of $L_1$ and $L_0$, we can get the points we
are interested on $L_0$ which correspond to $Q_i$. Since the
topology of the surface does not change on $L_1$ at $Q_2$, we need
not to consider the corresponding point on $L_0$. Let
$V_0,V_1,V_2,V_3,V_4$ be the points on $L_0$ corresponds to
$Q_0,Q_1,Q_3,Q_4,Q_5$ on $L_1$. We have the points
$$\begin{array}{rl}
&[V_0,V_1,V_2,V_3,V_4]=[[[0,0]\times[0,0]\times t],\\
&t=[-2,-2],[-\frac{7}{4}, -\frac{7}{4}],[0,0],[\frac{7}{4},
\frac{7}{4}],[2,2]]\end{array}$$ and the edges $(V_0,V_1),
(V_1,V_2), (V_2,V_3), (V_3,V_4)$.

Now we need to find out the SCCSes of the original surface which
originate from these points and edges on vertical line. The basic
idea is as below.

At first, find a separate point $W_i$ on each vertical edge, that
is, between two adjacent points $V_{i-1},V_{i}$,  then construct a
plane $\Theta_{W_i}$ paralleling to XY-plane passing $W_i$, search a
rectangle $R_i$ containing $W_i$ such that all the curve segments
inside $R_i$ originate from $W_i$, and when projected into
$XY$-plane, all this kind of $R_i$ correspond to a same rectangle
$R$ which only contains one critical point $P$. In order to
determine the number of SCCSes originating from each vertical point,
we need the following lemma.

\begin{lemma}
The number of SCCSes originating from the point $V_i$ equals the
number of intersections of line $\{x=\alpha,y=\beta\}$ and the
surface $\S$ between two planes $\Theta_{W_i}$ and
$\Theta_{W_{i+1}}$, where $(\alpha,\beta)$ is a point on $C(e)$
inside $R$.
\end{lemma}
{\bf Proof.} Since there is only one vertical points between
$\Theta_{W_i}$ and $\Theta_{W_{i+1}}$, the SCCSes between two planes
originate from $V_i$. There is no part of the surface in $R_i$ or
$R_{i+1}$ has intersection with $I(e)$ when projected to $R$.
Otherwise, there exists a critical point on $R$ besides $P$. By
Lemma \ref{lm-3c}, the SCCS originating from $V_i$ only intersects
the line $\{x=\alpha,y=\beta\}$ once. So the lemma is true. \qed

So we have the method to decide the number of SCCSes for vertical
line case.

Then, we need to decide the number of surface patches originating
from each vertical line segment in each plane cell. In fact, this is
done! The boundaries of $R_i$ have some intersections with the
surface, the number of the intersections in each cell body is the
number of surface patches originating from the corresponding
vertical line segment. For our example, since $V_0,V_4$ are
endpoints and $(V_0,V_1),(V_3,V_4)$ do not originate surface patch,
we just need to find rectangles for $(V_1,V_2), (V_2,V_3)$.
So we can conclude that
$(V_1,V_2)$ originate two surface patches in cell bodies ``2'' and
``4'' respectively, and $(V_2,V_3)$ originate two surface patches in
cell bodies ``1'' and ``3'' respectively. When computing the SCCSes
originating from $V_1,V_2,V_3$, we can find that $V_1,V_3$ do not
originate non-vertical SCCSes, $V_2$ originates four line segments
as SCCSes.

In the end, we should form triangles for this case. The curve
branches in $R_i$ can intersect the plane triangles when projected
to $XY$-plane. Use these points to subdivide the plane triangles,
and then to form triangular patches. Note that when an endpoint of a
plane triangle corresponding to a vertical line, some of the surface
patches corresponding to the triangle should contain two or three
TSPs.

Figure \ref{fig-vertical} is a triangular polyhedron representation
of the surface defined by Equation \ref{eq-vertical} which has a
vertical line $\{x=0,y=0\}$.

\section{Ambient isotopic meshing of surface}
\label{sec-sur-amesh}

In this section, we will show how to compute an $\epsilon$-meshing
of a surface $\S$ for a given $\epsilon>0$.

Let $\MM_1$ be an $\epsilon$-meshing graph of the projection curve
of $\S$ computed with Algorithm \ref{alg-atopcur}.
%
Consider the two disjoint regions of $\B_3$:
\begin{eqnarray}
\SB_3 &=& \cup_{e\in\MM_1} \B_e\times[\ZB_1,\ZB_2] \label{eq-s11}\\
\NB_3 &=& \B_3\setminus \SB_3.\label{eq-s12}
\end{eqnarray}

Surface $\S$ has no singularities in the cylindrical region $\NB_3$,
so we can use a modified Pantinga-Vegter method \cite{pv1} to
compute its meshing. What we need to do is to compute the correct
meshing inside $\SB_3$. To present the algorithm, we need
preparations given in Sections \ref{sec-ms1} and \ref{sec-ms2}.

\subsection{Extremal points of surfaces and spatial curves}
\label{sec-ms1}

In order to give an ambient isotopic meshing for a surface, 
we need to consider $z$-extremal points of surfaces and spatial
curves.
A point is called  {\bf $z$-extremal} if the surface achieves a
local extremum value at this point in the $z$-direction. We have

\begin{lemma}\label{lm-i3}
Let $f(x,y,z)=\prod_i f_i(x,y,z)$ be a square free polynomial and
$f_i$ irreducible polynomials. A necessary condition for  the
surface $f(x,y,z)=0$ to have a $z$-extremal point is
\begin{equation} \label{eq-ms1}
 G_1(x,y)= \prod_i \res(f_i,\frac{\partial f_i}{\partial x},z)
           \prod_i \res(f_i,\frac{\partial f_i}{\partial y},z)=0
 \end{equation}
where only the nonzero resultants are included.
\end{lemma}
The following example shows that we need to consider the irreducible
factors.
Let $f = (z-y)(z-x)(x^2+y^2+z^2-1)$. Then
$\res(f,f_x,z)=\res(f,f_y,z)\equiv0$. But the surface indeed has an
$z$-extremal point at $(0,0,1)$.

\begin{lemma}\label{lm-i4}
Let $f(x,y,z)$ be a square free polynomial, $D(x,y)$ defined in
\bref{eq-D}, $G_1(x,y)$ defined in \bref{eq-ms1}, and $r$ a fixed
number.  Then $D(x,y)G_1(x,y)=0$ is a necessary condition for the
curves $f(x,y,r)=0$, $f(x,r,z)=0$, and $f(r,y,z)=0$ to have
$x$-extremal, $y$-extremal, or $z$-extremal points.
\end{lemma}

We also need to consider the $z$-extremal points of spatial curves
defined by $g(x,y)= f(x,y,z)=0$, where $g$ and $f$ are polynomials.
For this purpose, we need to decompose the curve into irreducible
ones.
The leading coefficient of $g$ ($f$) as an univariate polynomial in
$y$ ($z$) is called the {\bf initial} of $g$ ($f$).
Two polynomials of the form $g(x,y), f(x,y,z)$ is called an {\bf
irreducible chain} if the following conditions are satisfied
\cite{mi} (pages, 297-381)
\begin{itemize}
\item $g(x,y)$ is an irreducible polynomial.
\item
$f(x,y,z)$ is an irreducible polynomial of $z$ module $g=0$,
$\deg(f,y) < \deg(g,y)$, and the initial of $f$ is a polynomial in
$x$.

\end{itemize}
For instance, $g=y^2-x,f=z^2-x$ is not irreducible, since $f =
(z-y)(z+y) + g = (z-y)(z+y) \mod(g)$.

For an irreducible chain $g(x,y), f(x,y,z)$, we define its {\bf
saturation ideal} to be
 $$\sat(g(x,y),f(x,y,z)) =\{P\,|\, I_1^sI_2^kP\in(f,g)\}$$
 where $I_1$ and $I_2$ are the initials of $g$ and $f$ respectively.
It is known that the saturation ideal of an irreducible chain is a
prime ideal, and thus defines an irreducible spatial curve \cite{mi}
(pages, 297-381).

Any spatial curve $f(x,y,z)=g(x,y)=0$  can be decomposed into the
union of irreducible curves algorithmically:
 \begin{equation}\label{eq-dec}
 V(g(x,y),f(x,y,z)) = \cup_i
 V(\sat(g_i(x,y),f_i(x,y,z)))
 \end{equation}
where $g_i(x,y),f_i(x,y,z)$ are irreducible chains.
We can prove the following result:
\begin{theorem} Let $g(x,y),f(x,y,z)$ be an irreducible chain
and  {\small \begin{eqnarray}\label{eq-ec}
 I(x)&=&\hbox{ product of the initials of }f,g.\\
 T(x)&=&\res(\res(h,f,z),g,y) \hbox{ where }
 h(x,y,z)=f_xg_y-f_yg_x.\nonumber
 \end{eqnarray}}
Let $E$  be the  set of $z$-extremal points of the curve $\C:
f=g=0$. Then
 \begin{equation}\label{eq-112}
 Proj_x(E)\subset V(T(x))\cup V(I(x)).\end{equation}
Furthermore, if $T(x)\equiv0$, then the curve $V(\sat(\{f,g\}))$ is
contained in several planes perpendicular to the $z$-axis.
\end{theorem}

{\em Proof. } For any point $P=(\alpha,\beta,\gamma)$ on $\C$, the
necessary condition of $P$ being a $z$-extremal point of $\C$ is the
tangent line of $\C$ at $P$ is perpendicular to  $z$-axes. If $P$ is
neither the singular point of $f=0$ nor $g=0$, the tangent planes of
$f=0$ and $g=0$ at $P$ are both well defined. The tangent line of
$\C$ at $P$ is the intersection of the tangent planes of $f=0$ and
$g=0$ at $P$. The tangent direction {\bf n} of $\C$ at $P$ is
$$\begin{array}{rcl}
{\bf
n}&=&<f_x,f_y,f_z>|_P\times<g_x,g_y,0>|_P\\
&=&<-g_yf_z,f_zg_x,f_xg_y-f_yg_x>|_P.\end{array}$$ Since the tangent
planes of $f=0$ and $g=0$ at $P$ are:
$$\small\left\{\begin{array}{l}
(x-\alpha)f_x(\alpha,\beta,\gamma)+(y-\beta)f_y(\alpha,\beta,\gamma)+(z-\gamma)f_z(\alpha,\beta,\gamma)=0,\\
(x-\alpha)g_x(\alpha,\beta)+(y-\beta)g_y(\alpha,\beta)=0
\end{array}\right.$$
respectively. ${\bf n}$ is perpendicular to the $z$-axes, that is:
$$\begin{array}{rcl}
{\bf n}\cdot<0,0,1>&=&f_x(\alpha,\beta,\gamma)g_y(\alpha,\beta)-f_y(\alpha,\beta,\gamma)g_x(\alpha,\beta)\\
&=&h(\alpha,\beta,\gamma)=0.\end{array}$$ Therefore, $E\subset
V(h(x,y,z))$. \bref{eq-112} is true. \vspace{5pt}

If $T(x)\equiv 0$, we prove that $V(\{f,g\}/I)$ is contained in
several planes perpendicular to the $z$-axis.

First we claim that ${\bf n}$ is well defined on $\C$ except finite
number of points, that is only finite number of points on $\C$ are
the singular  points of $f=0$ or $g=0$. If it is not true, at least
one of the following conditions occurs:
\begin{itemize}
\item[C1.] $V(f,g,f_x,f_y,f_z)$ has 1-dimensional component.
\item[C2.] $V(f,g,g_x,g_y)$ has 1-dimensional component.
\end{itemize}
If C1. occurs, it means that $f_z\in \sat(f,g)$. It is impossible.
Condition C2. could not take place for the same reason. Note that
$h(x_0,y_0,z_0)=0$ for any point $(x_0,y_0,z_0)$ on $\C$. The
tangent direction of  $\C$ at almost all points is the form
$(A,B,0)$.

Then we prove this component of $\C$ lies in some planes $z=z_0$.
This component of $\C$ can be parametrization in some segments.
Assume the parametric equation is ${\bf r}(t)$. We have
$${\bf r}(t)={\bf r}(t_0)+\int_0^t{{\bf r}'(t)}=(x(t),y(t),z_0),$$
where ${\bf r}(t_0)=(x_0,y_0,z_0)$. It implies that this segment of
$\C$ lies in the plane $z=z_0$. Therefore, the irreducible component
of $\C$ which contains this segment lies in the plane $z=z_0$.
We prove this theorem.\qed

The following example shows that we need to decompose the curve into
irreducible ones.
Let $f = z(x^2+z^2-1), g=y$. Then $\res(f_xg_y-f_yg_x,f,z)\equiv0$.
But the curve indeed has a $z$-extremal point at $(0,0,1)$.

\subsection{Compute segregating box for an SCCS} \label{sec-ms2}

In Section \ref{sec-edge}, we showed how to compute the segregating
box for a singular point. In this section, we introduce the concept
of segregating boxes for singular curve segments.

\begin{figure}[ht]
\begin{minipage}{0.34\textwidth}
\centering
\begin{center}
\includegraphics[scale=0.30]{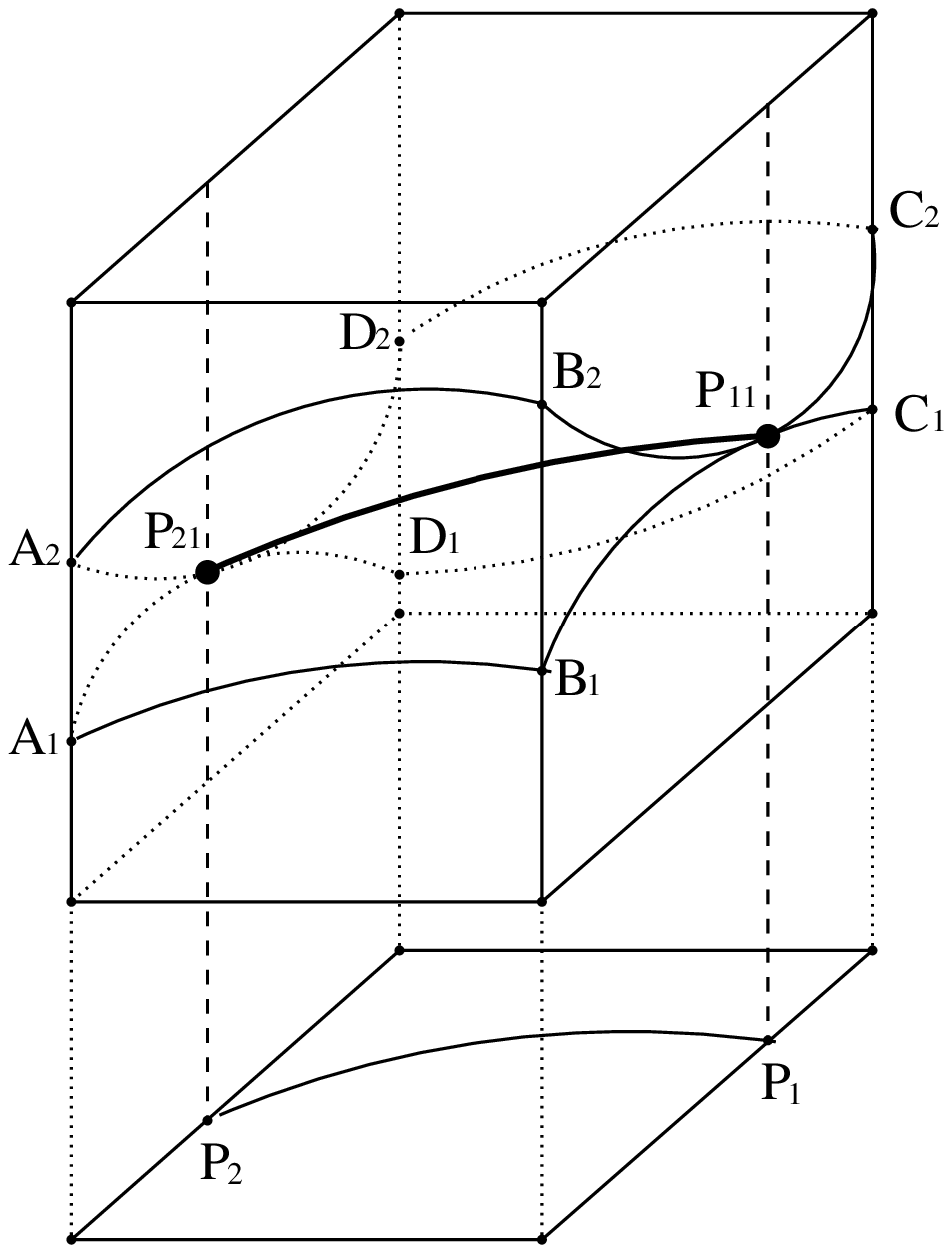}\quad
\caption{Segregating boxes} \label{fig-sbc}
\end{center}
\end{minipage}
\begin{minipage}{0.34\textwidth}
\includegraphics[scale=0.4]{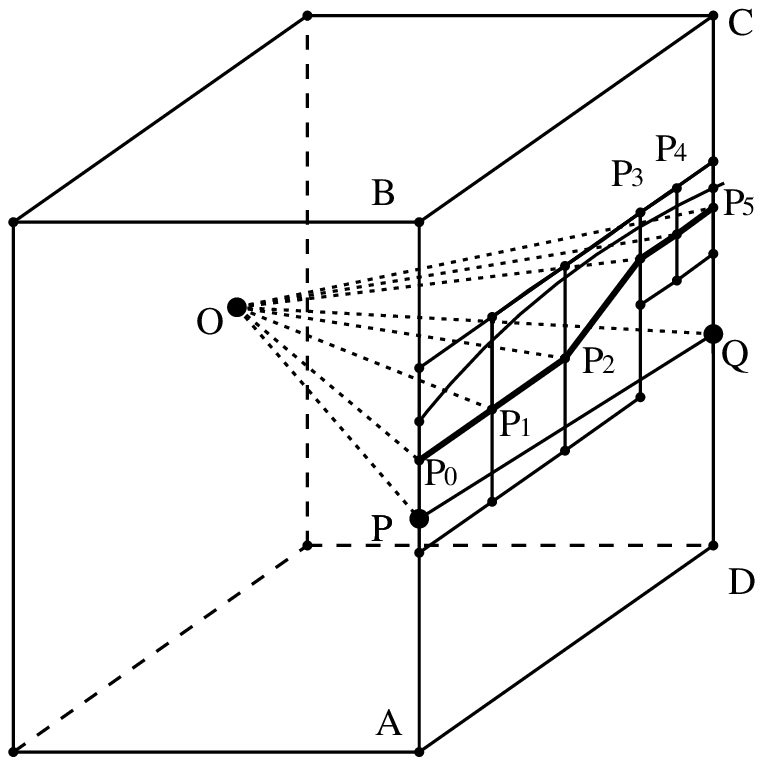}\quad

\vskip20pt \caption{Merge meshes} \label{fig-merge}
\end{minipage}
\begin{minipage}{0.30\textwidth}
\centering
\includegraphics[scale=0.25]{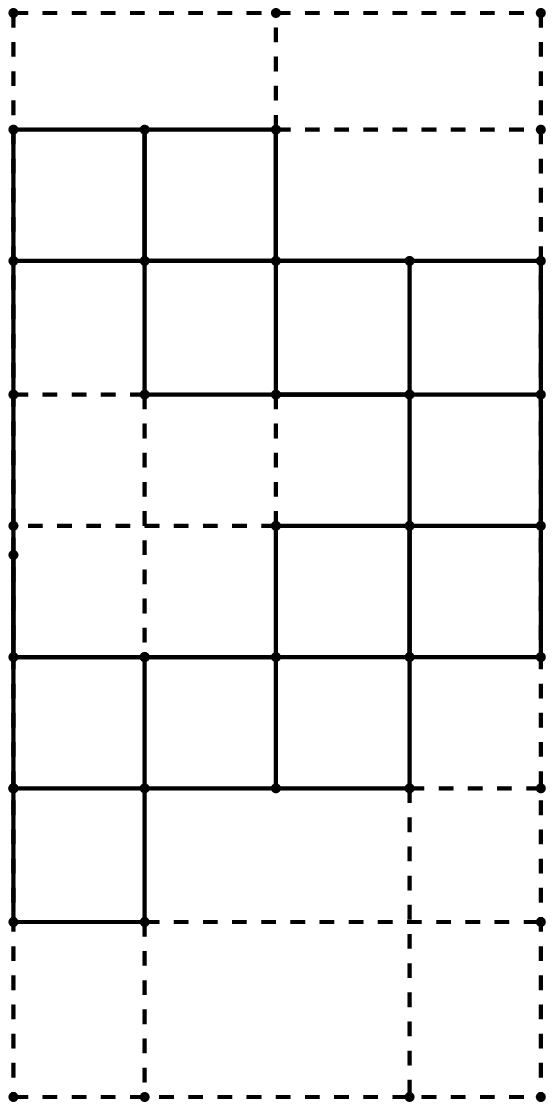}
\caption{Divide $\NB_2$ into boxes}\label{fig-mpv2}
\end{minipage}

\end{figure}

In Algorithm \ref{alg-atopcur}, a curve segment $C(e)$ of the
projection curve $\C$ is represented by a segment $e$ contained in a
box $\B_e$, as shown in Fig. \ref{fig-ms-m}. When lifting $\C$ to
the space, we obtain a set of SCCSes $S_i,i=1,\ldots,d$  of $\S$
represented by edges $E_i\in\SE$ (see Section \ref{sec-s2}).
A box $\B_{S_i}=\B_e\times[e_i,f_i]$ is called a {\bf segregating
box} for $S_i$ if $\B_{S_j}\cap\B_{S_i}=\emptyset$ for $i\neq j$ and
$\S$ does not intersect with the top and bottom faces of $\B_{S_i}$.
In Fig. \ref{fig-sbc}, we give a segregating box for the surface
patches $A_1B_1C_1D_1$ and $A_2B_2C_2D_2$ intersecting at curve
segment $P_{11}P_{21}$ which is lifted from curve segment $P_1P_2$.

Assume that all $S_i$ are monotonous in the direction of $z$, the
following algorithm shows how to compute segregating boxes for the
SCCSes: $S_i,i=1,\ldots,d$.

\begin{alg}\label{alg-segboxc}{\bf SegBoxC$(f(x,y,z),g(x,y), \B_3,\B_{e},\epsilon)$.}
Let $\S:f(x,y,z)=0$ be the surface, $\B_3$ the bounding box,
$\B_{e}$ a nice box (see Fig. \ref{fig-ms-c}) containing a curve
segment $C(e)$ of the projection curve $\C:g(x,y)=0$ of $\S$,
$\epsilon>0$.

The output is a pair $(\PB,\SB)$. $\PB$ is a set of
interior-disjoint boxes contained in $\B_{e}$, the union of which
contains $C(e)$. For each $\PB_i\in\PB$, there exist 3D boxes
$\SB_{i,j}\in\SB$ which are the segregating boxes for the SCCSes
lifted from $C(e)\cap\PB_i$.

\end{alg}
\begin{enumerate}
\item
We consider case (a) in Fig. \ref{fig-ms-c}. Other cases can be
treated similarly.
$C(e)$ divides $\B_{e}$ into two cells $c_1$ and $c_2$.

\item Let $\PB_1=\{\B_{e}\}$, $\PB=\emptyset$, $\SB=\emptyset$. Repeat
the following steps until $\PB_1=\emptyset$.
\begin{enumerate}
\item Let $\B=[a,b]\times[c,d]\in\PB_1$ and remove $\B$ from
$\PB_1$.
\item
Execute
{\bf{RootIsol}}$(\{g(a,y),f(a,y,z)\},[c,d]\times[\ZB_1,\ZB_2]$,
$\epsilon)$ to compute the points $P_{1,i},i=1,\ldots,N_1$ lifted
from $P_1$. See Fig. \ref{fig-sbc} for an illustration.
Let the isolation box for $P_{1,i}$ be
$\SB_{1,i}\times[e_{1,i},f_{1,i}]$.

\item Similarly, let $P_{2,i}, e_{2,i}, f_{2,i},
i=1,\ldots,N_2$ be the points lifted from $P_2$.
By Lemma \ref{lm-3c}, $N_1=N_2$.

\item  Let
 $\B_{i}=\B_e\times[\min\{e_{1,i},e_{2,i}\},\max\{f_{1,i},f_{2,i}\}], i=1,\ldots,N_1.$

\item If $|\B_{i}|<\epsilon$ for all $i$, add $\B_e$ to $\PB$ and add
$\B_{i}$ to $\SB$.

\item Otherwise,  subdivide $\B_e$ into four equal boxes and add the
boxes intersecting with $\C$ into $\PB_1$.

\end{enumerate}

\item
Repeat the following steps until all boxes in $\SB$ are segregating.
\begin{enumerate}
\item Let $\bar{\B}=\B\times[e,f]\in\SB$.

\item
If $0\not\in\intbox f(\B\times[e,e])$ and $0\not\in\intbox
f(\B\times[f,f])$, then $\bar{\B}$ is segregating, we do nothing for
$\B$.

\item
Otherwise, remove $\B$ from $\PB$ and $\bar{\B}$ from $\SB$.
Subdivide $\B$ into four equal boxes $\CB_1,\CB_2$, $\CB_3,\CB_4$,
add each $\CB_i$ intersecting with $\C$ into $\PB$, and add
$\CB_i\times[e,f]$ into $\SB$.

\end{enumerate}

\item Return $\PB$ and $\SB$.

\end{enumerate}

{\em Proof of correctness.} We need only prove the termination of
the algorithm. According to the assumption, all $S_i$ are monotonous
in the $z$ direction. So Step 2 terminates in a finite number of
steps.

At the beginning of Step {\bf 3}, for any $C(e)\subset
\B=[a,b]\times[c,d]\in \PB$ where
$e=(P_1,P_2),~P_1=(a,\alpha),P_2=(b,\beta)$, let $C(s_j)$  be a
curve segment lifted from $e$ and
$\B_i=[a,b]\times[c,d]\times[e_i,f_i]$ be the corresponding box
where $s_i=(P_{1,i},P_{2,i}),~P_{1,i}=(a,\alpha,\gamma_i),
P_{2,i}=(b,\beta,\tau_i)$, we have $|\B_i|<\epsilon$ and
$$f(a,\alpha, e_i)f(a,\alpha,f_i)f(b,\beta,e_i)f(b,\beta,f_i)\neq
0.$$ Furthermore, $s_i$ does not intersect with the top nor bottom
faces of $\B_i$ since $s_i$ is monotonous in the direction of $z$.
That is $0\not\in \intbox f(C(e)\times [e_i,e_i])f(C(e)\times
[f_i,f_i])$. So there exists a positive number $\delta$ such that
$$0\not\in \intbox f(d(C(e),\delta)\times
[e_i,e_i])f(d(C(e),\delta)\times [f_i,f_i])$$ where $d(C(e),\delta)$
is a zonal region in $\R^2$ containing the points $Q$ such that the
distance between $Q$ and  $C(e)$ is less than $\delta$. We can get
the set of sub-boxes of $\B$ in a finite steps such that all boxes
in it are contained in the region $d(C(e),\delta)$. Then the
algorithm clearly terminates. \qed

\subsection{Compute $\epsilon$-meshing of surface}

Similar to the case of curves, we need to modify the Pantinga-Vegter
method.
A box $\B$ is called a {\bf nice box} if each face of $\B$ is a nice
2D box. For an illustration, see the 2D case in Fig. \ref{fig-ms-c}.
To make the process precise, we introduce the following definition.

A {\bf meshing polyhedron} of a surface $\S$ is a four-tuple
$\MM=\{\PP,\PE,\PF,\MB\}$ where
$(\PP,\PE,\PF)$ is a polyhedron whose vertices are with rational
numbers as coordinates and whose {\bf faces are the meshes} for
$\S$;
$\MB$ is a set of nice boxes and segregating boxes of singular
points of $\S$ s.t. for each $F\in\PF$, there exists a $\B_F\in\MB$
with the property that the surface patch $\S\cap\B_F$ is connected.

A meshing polyhedron $\MM$ is called an {\bf $\epsilon$-meshing
polyhedron} if each box $\B$ in $\MB$ satisfies  $|\B|<\epsilon$.
It is easy to show that an $\epsilon$-meshing polyhedron for a
surface $\S$ provides an {\bf $\epsilon$-meshing} for $\S$ according
to the definition given in Section 2.

\begin{alg}\label{alg-mpv3}
{\bf MPV3}$(f(x,y,z),\NB_3,\epsilon).$ $\S: f(x,y,z)=0$ is the
surface. $\NB_3$ is a box contains no zero of $D(x,y)G_1(x,y)=0$
where $D(x,y)$ is defined in \bref{eq-D} and $G_1(x,y)$ is defined
in \bref{eq-ms1}.
Output an $\epsilon$-meshing polyhedron for $\S_{\NB_3}$.
\end{alg}

\begin{enumerate}
\item 
Subdivide $\NB_3$ into boxes $\B_i$  at the corner lines (Fig.
\ref{fig-mpv2} shows how to subdivide the region $\NB_{2}^{1}$ in
Fig. \ref{fig-ms-ce}(a), where the dotted lines are newly added.)
and execute the Pantinga-Vegter algorithm with initial values
$\{\B_i\}$. Let $\SB$ be the output.

\item For each cube $\B\in\SB$, repeat subdividing $\B$ until all of the following
statements are false.
\begin{enumerate}
\item
There exists an edge $(A,B)$ of $\B$ s.t. $0\in \intbox f((A,B))$
and $f(A)f(B)>0$.

\item There exists a face $ABCD$ of $\B$ s.t.
$f(A)f(B)<0\wedge f(B)f(C)<0\wedge f(C)f(D)<0\wedge f(D)f(A)<0$.

\item $|\B|>\epsilon$.
\end{enumerate}
\end{enumerate}
Termination of the algorithm is guaranteed by Lemma \ref{lm-i4}.

Now we can compute the $\epsilon$-meshing  for $\S_{\mathbf{\B_3}}$.

\begin{alg}\label{alg-atopsur}
{\bf ATopSur}$(f(x,y,z),\B_3,\epsilon)$. The input is the same as
Algorithm \ref{alg-topsur}. The output is an {\bf $\epsilon$-meshing
polyhedron} for $\S_{\mathbf{\B_3}}$.
\end{alg}
\begin{description}


\item[S1]
{\bf Compute the critical points of the projection curve and their
segregating boxes.}

\begin{enumerate}

\item Let
\vspace{-5pt}
\begin{equation}\label{eq-ms3}
 G(x,y)=\hbox{sqrfree}(G(x,y) G_1(x,y)),
\end{equation}
where $G$ is defined in \bref{eq-projcur} and $G_1$ is defined in
\bref{eq-ms1}.

\item Execute the first four steps of Algorithm \ref{alg-topcur} with
input ($G(x,y),\B_2,\epsilon)$ to compute a set of points $\PP_1$
and the segregating box for each point in $\PP_1$.
We need to modify the algorithm as follows. In Step 3 of Algorithm
\ref{alg-topcur}, we use the new projection polynomial:
\vspace{-5pt}
\begin{equation}
 H(x):= H(x)\prod_i I_i(x)\prod_iT_i(x),
\vspace{-5pt}
\end{equation}
where $H(x)$ is defined in \bref{eqn-hdefinition}, $I_i(x)$ and
$T_i(x)$ are defined in \bref{eq-ec} with decomposition
\bref{eq-dec}. Only the nonzero $T_i$ are considered.

\end{enumerate}

\item[S2] {\bf Compute $\SP_0$ and the set $\SBX_0$ of segregating
boxes for points in $\SP_0$.} 
For any $P_{i,j}\in\PP_1$, use Algorithm \ref{alg-segboxp3} with
input $(f,\B_3,P_{i,j},\epsilon)$ to compute the points lifted from
$P_{i,j}$ and their segregating boxes.
Let $\MB_1$ be the set of all updated segregating boxes $\B_{i,j}$
of $P_{i,j}$. Let $\M_1=\{\PP_1,\MB_1\}$.

\item[S3]
{\bf Compute an $\epsilon$-meshing graph for the non-singular part
of $\C_{B_2}$ in $\NB_2$ defined in \bref{eq-i1}.}
Let $\MM_0=$ {\bf MPV2}$(G(x,y),\NB_2,\epsilon)$, where $\NB_2$ is
defined in \bref{eq-i1}.

\item[S4]{\bf  Compute segregating boxes for SCCSs:}

\begin{enumerate}

\item
Assume $\MM_0=\{\PP_0,\PE_0,\MB_0\}$. Let
$\SBX_1=\PP_2=\PE_2=\MB_2=\emptyset$.

\item For each $\B\in\MB_0$, execute the following steps:
\begin{enumerate}

\item Compute $\{{\bf P},\SB\}$={\bf
SegBoxC}$(f(x,y,z),G(x,y),$ $\B_3,\B,\epsilon)$.\footnote{Step {S1}
ensures all $z$-extremal points of the curve $\C: f=0,G=0$ are in
$\SB_3$. Hence the SCCS in $\B$ is monotonous in the direction of
$z$.}

\item $\SBX_1 = \SBX_1\cup \SB$ and update $\PP_2,\PE_2,\MB_2$ according to ${\bf P}$ which subdivides
$\B$.
\end{enumerate}

\item Let $\MM_2=\{\PP_2,\PE_2,\MB_2\}$.
\end{enumerate}

\item[S5]{\bf Compute the extended meshing graph $\EG_S$ of $\C$} with
Algorithm \ref{alg-emeshcur} with input  $\MM_1$ and $\MM_2$.

\item[S6] {\bf Meshing  the singular part of $\S$ in
$\SB_3$}.
\begin{enumerate}
\item
Let $\{\SP_1,\SE_1,\SF_1\}$={\bf TopSur}$(f(x,y,z),\B_3)$.
Modify Algorithm {\bf TopSur} as follows: use $G(x,y)$ defined in
\bref{eq-ms3} in Step 1, use $\EG_S$ in the Step 2, and use $\SP_0$
in Step 3. We actually only run Steps 4 and 5 of Algorithm {\bf
TopSur}.

\item
Let $\MM_1=\{\SP_1,\SE_1,\SF_1,\SBX_0\cup\SBX_1\}$ where $\SBX_0$
and $\SBX_1$  are from Steps S2 and S4 respectively.
$\MM_1$ is an $\epsilon$-meshing polyhedron for $\S_{\SB_3}$.

\end{enumerate}
\item[S7] {\bf Meshing the non-singular part of $\S$ in $\NB_3$.}
%
%
Let $\MM_2=\{\mp_2,\me_2,\mf_2,\mb_2\}=\mbox{\bf
MPV3}(f(x,y,z),\NB_3,$ $\epsilon)$, where $\NB_3$ is defined in
\bref{eq-s12}.

\item[S8]
{\bf Merge  $\MM_1$ and $\MM_2$ to obtain an $\epsilon$-meshing
polyhedron for $\S$.} Output $\mbox{\bf Merge}(\MM_1,\MM_2)$ (with
Algorithm \ref{alg-consur}).

\end{description}

\begin{theorem}\label{th-m2}
Algorithm \ref{alg-atopsur} computes an $\epsilon$-AIMESH for
$\S_{\mathbf{\B_3}}$.
\end{theorem}

{\em Proof. } The prove is similar to the  proof of Theorem
\ref{th-m1}. \qed

 In principle, there exist no difficulties to
implement the algorithm. But, it will take a lots of time, since we
need to incorporate algorithms from symbolic computation, interval
arithmetics, and marching cube into one program. This will be our
further work,


In the final step of Algorithm \ref{alg-atopsur}, we need to merge
two meshing polyhedrons, which will be done by the following
algorithm.
\begin{alg}\label{alg-consur}
{\bf Merge}$(\M_1,\M_2)$. $\M_1 =\{\Mp_1,\me_1, \mf_1$, $\mb_1\}$
and $\M_2 =\{\Mp_2,\me_2$, $\mf_2,\mb_2\}$ are the
$\epsilon$-meshing polyhedrons of $\S$ in ${\bf S}_3$ and ${\bf
N}_3$ respectively. The algorithm merges $\M_1$ and $\M_2$ and
outputs an $\epsilon$-meshing polyhedron $\M=\{\Mp,\me,\mf\}$ for
the surface.
\end{alg}

\begin{description}

\item[S1] Let $\mb_t=\mb_1$.

\item[S2] While $\mb_t\neq \emptyset$, repeat
\begin{enumerate}
\item Remove $\B=[a,b]\times[c,d]\times[e,f]$ from $\mb_t$.
Insert box $\B_i=[b,b_i]\times[c_i,d_i]\times[e_i,f_i]\in \mb_2$
which is connected with $\B$ according to $\S$ and adjacent to the
face ${\bf F}=[b,b]\times[c,d]\times[e,f]$  into $\B_a$ and insert
corresponding $V(\B_i)$ into $\PP_a$ . Pick out boxes $\B_i$
satisfying $d_i=\min_{\B_j\in \B_a}\{d_j\}$. Rename them to be
$\B_1,\ldots,\B_m$.  Sort the residual boxes in $\B_a$ as
$\{\B_{m+1},\ldots,\B_r\}$ such that
 $c_{m+1}\leq c_{m+2} \leq \ldots\leq c_r$ and for each $\B_k,k>m$,
$\B_k$ is connected with some $\B_j,~j<m$ according to $\S$(Note
that the result is not unique, and any $B_k,0<k<m$ only overlaps
with $\B$ on the vertical edge
$[b,b]\times[d_i,d_i]\times[e_i,f_i]$).

\item For $i$ from 1 to $r$ do

\begin{enumerate}
\item[(a)] Remove the points $P$ from $V(\B_i)$  and insert points
$Q\in\SP\cap(\B\cap\B_i)$ into $V(\B_i)$ if $P\in {\bf L}\cup{\bf
R}$ where ${\bf L}=[b,b]\times[c,c]\times[e,f],~ {\bf
R}=[b,b]\times[d,d]\times[e,f]$. Remove edge $(P,P_i)$ from $\me_2$
which are the edges with $P$ as an ending point and insert $(Q,P_i)$
into $\mf_2$. Remove triangular faces $(P,P_i,P_j)$ from $\mf_2$
which are the faces with $(P,P_i)$ as an edge  and insert
$(Q,P_i,P_j)$ into $\mf_2$.

\item[(b)] If there exist $P\in {\bf R_i}$ where ${\bf R_i}=[b,b]\times[d_i,d_i]\times[e_i,e_i]$
and ${ \bf R_i}\cap{\bf L}\cap{\bf R}=\emptyset$, add $P$ into
$\Mp_1$. goto (e).

\item[(c)] If there exist$P\in {\bf D_i}\subset {\bf F}$ where ${\bf D_i}=[b,b]\times[c_i,d_i]\times[e_i,e_i]$. Add $P$
in $\SP$. goto (e).

\item[(d)] If there exist$P\in {\bf U_i}\subset {\bf F}$ where ${\bf U_i}=[b,b]\times[c_i,d_i]\times[f_i,f_i]$. Add $P$
in $\SP$. goto (e).

\item[(e)]  Assume the
other point contained in the face
$[b,b]\times[c_i,d_i]\times[e_i,f_i]$ of $\B_i$ is $Q$ and
$(Q,S,T)\in\mf_1$ is the triangular face with $Q$ as a vertex where
$S\in {\bf R}$. Remove $(Q,S)$ from $\me_1$  and insert
$(Q,P),(P,S)$ into $\me_1$. Remove triangular faces $(Q,S,T)$ from
$\mf_1$ and insert $(Q,P,S),~(P,S,T)$ into $\mf_1$(The four edges of
this face of $\B_i$ contains two points. We can always assume that
we have dealt with the other one point, since $\B_i,i>m$ is
connected with some $\B_j$ we have dealt with).

\item[(f)] Update $\mb_2$ according to the new $V(\B_i)$.

\end{enumerate}

\item Determine the connection information of the other three faces
of $\B$ in the similar way.

\end{enumerate}

\item[S3] Out $\M=\{\Mp,\me,\mf\}$ where $\Mp=\Mp_1\cup\Mp_2,\me=\me_1\cup\me_2$, and
$\mf=\mf_1\cup\mf_2$.

\end{description}

A box $\B_1$ is said to be {\bf adjacent to a box} $\B_2$ \wrt the
surface $\S$ if $\B_1$ and $\B_2$ are interiorly disjoint and $\S$
intersects $\B_1\cap\B_2$. We need only consider how to merge the
meshes in two adjacent boxes.

We use the example in Fig. \ref{fig-merge} to explain the algorithm.
The large box $\B$ is in $\SB_3$ and contains singularities. We
consider the right face $\FB$ of $\B$.
Let $\B_i, i=1,\ldots,5$ be the boxes in $\NB_3$ adjacent to $\B$ at
face $\FB$. By Step 1 of Algorithm \ref{alg-mpv3}, $\B_i$ must be
completely between lines $AB$ and $CD$.
We will adjust the meshes in $\B$ and leave the meshes in $\B_i$
unchanged.
Since all the meshes are triangular, let $OPQ$ be the mesh of $\S$
in $\B$, and $N_iP_{i-1}P_{i}$ the mesh of $\S$ in $\B_i$. We will
replace the mesh $OPQ$ with the meshes $M_i =OP_{i-1}P_{i},
i=0,\ldots,4$.
If $P_i$ is above $BC$, $P_i$ is taken to be the intersection of
$BC$ and the line passing through $P_i$ and parallel to $AB$.
Other cases can be treated similarly.

\section{Conclusion}

This paper proposes complete methods to compute isotopic and ambient
isotopic meshings for implicit algebraic curves and surfaces. We use
symbolic computation to achieve completeness and whenever possible
use interval arithmetics to achieve practical effectiveness. Note
that an isotopic meshing without precision and an $\epsilon$-meshing
are quite different and can be used for different purposes.


\end{document}